\documentclass[12pt,preprint]{aastex}

\slugcomment{to be published in the Astrophys. J. 694 (April 2009)}

\shorttitle{21 Micron Emission in PPNs}
\shortauthors{Hrivnak, Volk, \& Kwok}

\begin{document}
\title{A {\it Spitzer} Study of 21 and 30 Micron Emission in Several Galactic Carbon-rich Proto-Planetary Nebulae}

\author{Bruce J. Hrivnak}
\affil{Department of Physics and Astronomy, Valparaiso University,
Valparaiso,IN 46383; bruce.hrivnak@valpo.edu} \email{}

\author{Kevin Volk}
\affil{Gemini Observatory, 670 North A`ohoku Place, Hilo, HI
96720; kvolk@gemini.edu} \email{}

\and

\author{Sun Kwok}
\affil{Department of Physics, The University of Hong Kong, Hong Kong, China; and \\
Department of Physics \& Astronomy, University of Calgary,
Calgary, Alberta, Canada T2N 1N4; sunkwok@hku.hk} \email{}

\begin{abstract}

We have carried out mid-infrared spectroscopy of seven Galactic
proto-planetary nebulae (PPNs) using the {\it Spitzer Space Telescope}.
They were observed from 10$-$36 $\mu$m at relatively high spectral
resolution, R$\approx$600.  The sample was chosen because they all
gave some evidence in the visible of a carbon-rich chemistry.  
All seven of the sources 
show the broad, unidentified 21 $\mu$m emission feature; 
three of them are new detections (IRAS 06530$-$0213, 07430+1115, 
and 19477+2401) and the others are observed at higher
S/N than in previous spectra. These have the same shape and
central wavelength (20.1 $\mu$m) as found in the {\it ISO} spectra
of the brighter PPNs.  The 30 $\mu$m feature was seen in all seven
objects. However, it is not resolved into two separate features
(26 and 33 $\mu$m) as was claimed on the basis of {\it ISO}
spectra, which presumably suffered from the noisy detector bands
in this region. All showed the infrared aromatic
bands (AIB) at 11.3, 12.4, and 13.3 $\mu$m. Five of these also
appear to have the C$_2$H$_2$ molecular band at 13.7 $\mu$m, one in
absorption and four in emission. 
This is extremely rare, with only one other evolved star, IRC$+$10216, 
in which C$_2$H$_2$ emission has been observed.
Four also possessed a broad, unidentified emission feature at 15.8 $\mu$m 
that may possibly be related to the 21 $\mu$m feature. 
Model fits were made to the spectral energy distributions for these PPNs to 
determine properties of the detached circumstellar envelopes.
The 21 $\mu$m feature has been seen in all Galactic carbon-rich PPNs observed, and thus its carrier appears to be a common component of the outflow around these objects.
\end{abstract}

\keywords{circumstellar matter: --- infrared: stars --- ISM:
planetary nebulae: general -- radiative transfer --- stars: AGB
and post-AGB }

\section{INTRODUCTION}


The study of infrared emission features in evolved stars provides an 
important  avenue to study the chemistry of the circumstellar environment.
First seen in the {\it IRAS} spectra of four sources \citep{kwo89}, 
a new, broad feature was discovered at $\sim$21 $\mu$m 
(the so-called ``21 $\mu$m feature'').  
Subsequently, a small number of additional sources with this feature have been
identified based on spectra from {\it IRAS}, {\it ISO}, and with
the CGS on the United Kingdom Infrared Telescope.  The total
number of such sources is about a dozen \citep{kwo99a}.  An {\it
ISO} study of the strongest sources showed that they all had the
same peak wavelength, 20.1 $\mu$m, and nearly the same shape with
no evidence of substructure \citep{vol99}.

The 21 $\mu$m sources all possess two characteristics in common.
First, the 21 $\mu$m sources are all carbon rich.  This was first deduced based 
on the presence of molecular C$_2$ and C$_3$ in their visible spectra
\citep{hri91a,hri95,bak97} and later confirmed in detailed
high-resolution abundance studies \citep{win00,red02}.
The second common characteristic of the 21 $\mu$m sources is that 
they are nearly all proto-planetary nebulae (PPNs), objects
in transition between the asymptotic giant branch (AGB) and
planetary nebula (PN) stages of the evolution of intermediate-mass
stars. During this stage, the extensive AGB mass loss has ceased
and the central star is surrounded by a detached, expanding
envelope of molecular gas and dust. This stage can last several
thousands of years \citep{blo95b}, during which the central star
evolves from spectral type late-G to early-B, after which time it
becomes hot enough to photoionize the gas and produce a PN.  
\citep[The 21 $\mu$m feature also appears to be present but weak in a few
young PNs;][]{hon01,vol03}.

There has been an active search for the carrier of the 21 $\mu$m
feature and many candidates have been suggested, including PAHs,
fullerenes, and nanodiamonds.  Most but not all of the proposed
carriers involve carbon.  A good match to the feature is produced
by TiC \citep{von00}, but this requires severe formation
constraints due to the low abundance of Ti, even in stars like
these that have enhanced s-process elements.  SiC also provides a
fit in this region and without the severe abundance constraints
(see \citet{spe04} and references therein).  However, most 
carriers suggested have associated emission features at other 
wavelengths that are absent or unconfirmed in these sources.

A very broad, so-called ``30 $\mu$m feature'' has been know for
several decades to exist in the spectra of evolved C-rich AGB
stars and PNs \citep{for81}.  It has been attributed to MgS
\citep{goe85}, which peaks in the range of 30$-$35 $\mu$m
depending on the temperature of the grain.  Subsequent studies
have supported this identification \citep{szc99,hon02}. It has
been seen with {\it ISO} in a number of objects ranging from AGB
stars to PPNs to PNs \citep{hon02, vol02} on a dust continuum that
peaks at monotonically longer wavelengths in the transition
between these stages. On the basis of recent {\it ISO} spectra, it
has been claimed that the feature has been resolved into two
features centered at 26 and 33 $\mu$m in PPNs, PNs, and
extreme-AGB stars \citep{hri00, vol02}.

A family of emission features at 3.3, 6.2, 7.7, 11.3, 12.4, and
13.3 $\mu$m (historically known as unidentified infrared [UIR]
bands) is commonly observed in the diffuse ISM. Although the
specific carriers are not identified, it is quite certain that
these arise from the stretching and bending modes of carbonaceous
materials with an aromatic component, such as polycyclic aromatic
hydrocarbons (PAHs). We will refer
to these as aromatic infrared bands (AIBs).  Since these AIBs are
also observed in carbon-rich PNs and PPNs, the carriers of these
bands are likely to be synthesized in the circumstellar
environment of late-type stars.  These are often accompanied in
PPNs by features at 3.4 and 6.9 $\mu$m attributed to aliphatic
groups attached to the aromatic hydrocarbons \citep{hri07}.  The
11.3, 12.4, and 13.3 $\mu$m features lie within the wavelength
range of the {\it Spitzer} spectrographs.


In this study, we have obtained {\it Spitzer} spectra of seven
PPNs to further study the 21 $\mu$m, 30 $\mu$m, and AIB features.
These targets were selected with the goal of completing the mid-infrared 
study of carbon-rich PPNs.  In addition to new targets that met this criterion (IRAS 06530$-$0213, 07430+1115, 19477+2401), we included some carbon-rich PPNs known or suspected to possess the 21 $\mu$m feature based on low S/N {\it ISO} or ground-based spectra.  
Also included for comparison was one additional object that was well 
observed with {\it ISO} but faint enough for {Spitzer} observations.  

\section{OBSERVATION AND DATA REDUCTION}

The target objects were observed with the Infrared Spectrograph
\citep[IRS;][]{hou04} on the {\it Spitzer Space Telescope}
\citep[{\it Spitzer};][]{wer04}.  Most were observed as part of
program No. 20208, PI: B. Hrivnak.  The targets were all
relatively bright for {\it Sptizer} and consequently were all
observed in the high resolution modes, Short-High (SH) and
Long-High (LH).  These modes consisted of ten overlapping orders,
and ranged in wavelength from 10$-$19.6 $\mu$m for SH to 19$-$37
$\mu$m for LH.  The slit sizes for the two modes are 4.7$\arcsec$
$\times$ 11.3$\arcsec$ for SH and 11.1$\arcsec$ $\times$
22.3$\arcsec$ for LH.  Each observation consists of two nod
positions, located at one-third and two-thirds of the way across
the slit; at each nod position three scan cycles were carried out.
The spectral resolution is R $\approx$ 600. No off-source
background measurements were made for these bright targets.  An
observing log is given in Table \ref{obs_log}.

\placetable{obs_log}

Two of the targets were initially observed using inaccurate coordinates and 
were re-observed during Director's Discretionary time.  
Three of the objects were not observed in this program in the SH
mode because they were on the GTO program no. 93, PI: D.
Cruikshank. Data for these were obtained from the {\it Spitzer}
archive and have been incorporated into this study.  Their
observations were made in a similar manner to those described
above.  Details of these additional observations have also been
included in Table \ref{obs_log}.

Unfortunately, the coordinates of some of the targets were not
know accurately and thus the stars were not always centered well
on the slits.  Accurate positions are important since the slits are
relatively small compared to those used in previous mid-infrared spectral
observations.  In some cases, this meant that the spectrum at one
or both of the two nod positions was weak or noisy or could not be
extracted properly. Peak up was performed with nearby stars using
the PCRS.  To avoid such problems in follow-up studies, accurate
positions are list in Table \ref{pos} for each of the targets as
derived from accurate ($<$1$\arcsec$) 2MASS positions. For the
fainter visible sources, source identifications are based on
finding charts determined from our ground-based mid-infrared
observations.  Finding charts for IRAS 19477+2411\footnote{Based
on comparison with the 2MASS catalog, it appears that the JHKL$\arcmin$
magnitudes that we published for IRAS 19477+2401 \citep{su01} were
for the nearby source located 8$\arcsec$ west and 5$\arcsec$
south, 2MASS 19495429+2408481.  The 2MASS magnitudes for IRAS
19447+2401 are J=12.61, H=10.75, and K=9.61.} and 22574+6609 are
shown by \citet{hri85} and \citet{hri91b}, respectively.

\placetable{pos}

The data were initially processed using the data reduction
pipeline at the Spitzer Science Center (SSC): version 13.2.0 for the
earlier data and version 16.1.0 for the two that were re-observed. 
We then used the software program SMART \citep{hig04} to reduce the
data. We started with the Basic Calibration Data (BCD) files.
These were first cleaned for bad pixels using IRSclean. Bad pixels
included the so-called rogue pixels, which are identified by
observing campaign at the SSC, and other bad data which are
flagged in the pipe-line. The spectra were then extracted and
defringed.  The resulting spectra were examined manually, by
order, within each of the individual scan cycles. Bad data were
removed from the edges of the orders. Data spikes not present in
all the individual scan cycles or in both of the overlap regions
in adjacent orders or in both of the nod positions were treated as
artifacts, as were some additional large spikes, and these were
removed.  An average spectrum was composed in each nod position.
In some cases we made small shifts in flux (2$-$10$\%$) to bring
the overlapping orders into better agreement.  The spectra from
the two nod positions were then compared and an average spectrum
formed if the two spectra agreed. In some cases, as mentioned
above, the spectrum from one of the nod positions was near the
edge of the slit and the data were of inferior quality; these
spectra were not included.  The spectral region 14.5$-$14.6 $\mu$m
had lots of spikes in the data and was not a region with
overlapping orders; this resulted in particularly noisy spectra in
this region.  The LH spectra beyond 29 $\mu$m were more noisy and
those beyond 36 $\mu$m were completely deleted.

The SH and LH spectra for each object were then combined, scaling
where necessary based on a comparison of the region of spectral
overlap from 19.0 to 19.6 $\mu$m.  
In all cases the SH was fainter or similar to the LH;
this may be due to the larger slit size of the LH mode, better accommodating 
small pointing errors.  
Therefore we scaled the SH when necessary to match the LH in
the overlap region to produce a final Spitzer spectrum for reach
object covering the region 10 to 36 $\mu$m.

We note for comparison purposes that previous
mid-infrared spectra have been obtained of some of the target objects.  
Initial observations were at low resolution
from space with LRS (7.7$-$23 $\mu$m, R$\sim$25) on {\it IRAS}, or
from the air using the Kuiper Airborne Observatory (KAO; 16$-$48
$\mu$m, R$\sim$33), or from the ground using CGS3 (7.5$-$13
$\mu$m, R$\sim$50; 16$-$24 $\mu$m, R$\sim$72) on the 3.8-m United
Kingdom Infrared Telescope (UKIRT). Higher resolution observations
were made more recently with SWS01 (R$\sim$250) or SWS06
(R$\sim$600$-$2000) on {\it ISO}.  
Some of these will be referenced in our following discussion of the new 
higher-resolution, higher signal-to-noise ratio {\it Spitzer} spectra.

\section{THE SPITZER IRS SPECTRA}

\begin{figure}
 \plotone{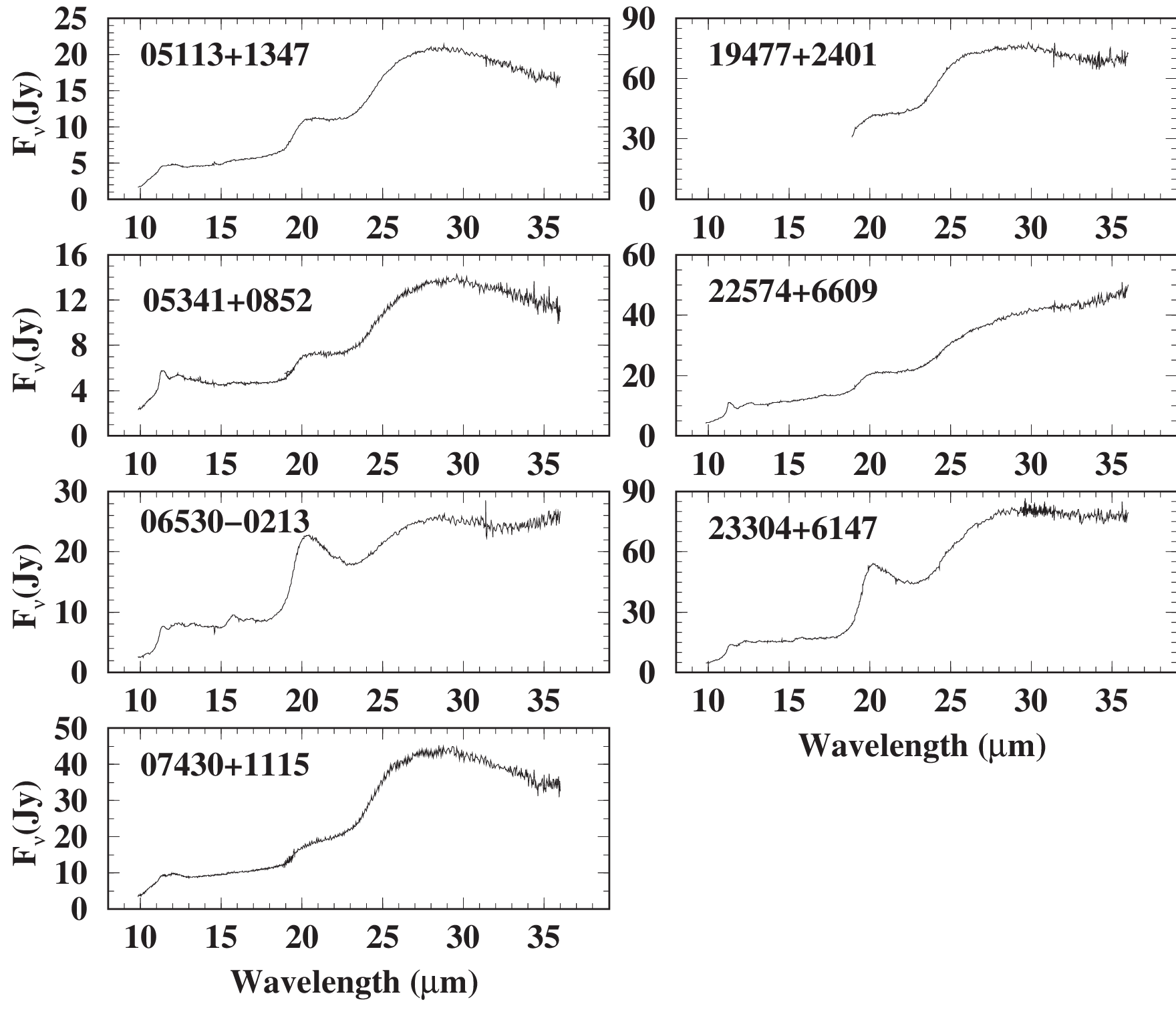}\figcaption[] {{\it Spitzer} IRS
high-resolution spectra of the seven sources observed with LH.
They each display the 21 $\mu$m emission feature. \label{fig1}}
\end{figure}

\begin{figure} 
\plotone{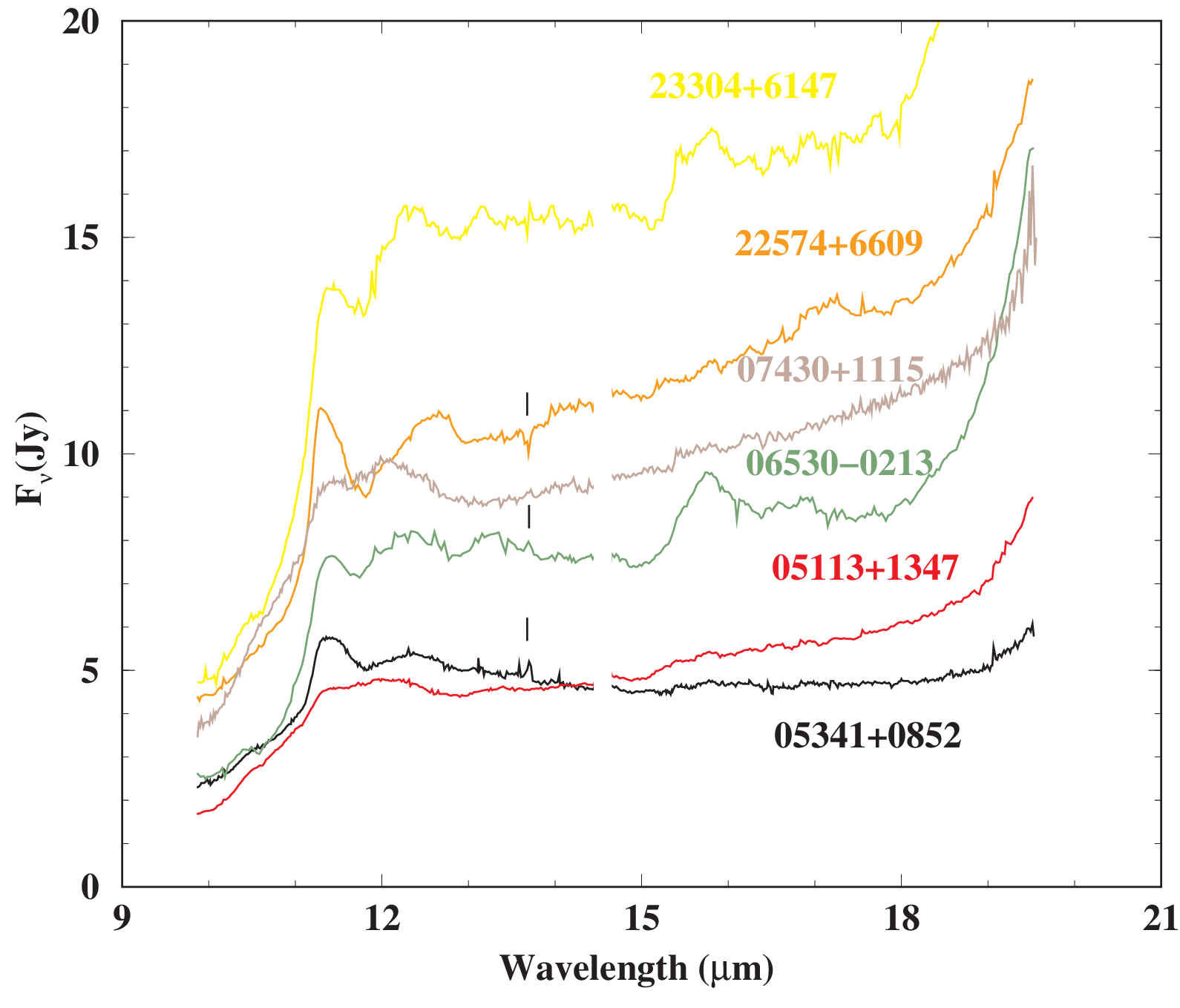}\figcaption[] {{\it Spitzer} IRS
high-resolution SH spectra of the six sources observed in this
mode.  They all show AIB emission features. The feature at 13.7 $\mu$m
(attributed to C$_2$H$_2$) seen in three of the spectra is indicated; 
it may also be present in IRAS 23304+6147 and 07430+1115.  
(The noisy 14.5-14.6 $\mu$m spectral region has been
removed from the spectra). \label{fig2}}
\end{figure}

{\it IRAS 23304+6147}: The SH and LH spectra in the two nod
positions were in good agreement and were averaged.  In the LH,
the longest wavelength order (LH11: 34$-$36 $\mu$m) was scaled
upward by 5$\%$. The complete spectrum is shown in Figure
\ref{fig1}. The spectrum shows strong broad 21 $\mu$m and 30 $\mu$m
emission features. Also seen are medium-width emission features at 11.4, 12.3,
13.2 $\mu$m, a medium-width feature at 15.9 $\mu$m and a weak feature at
22.3 $\mu$m. The short wavelength SH region is shown separately in
Figure \ref{fig2}, where the features in that wavelength region
can be seen more clearly.

\placefigure{fig1}

\placefigure{fig2}

\begin{figure}
\plotone{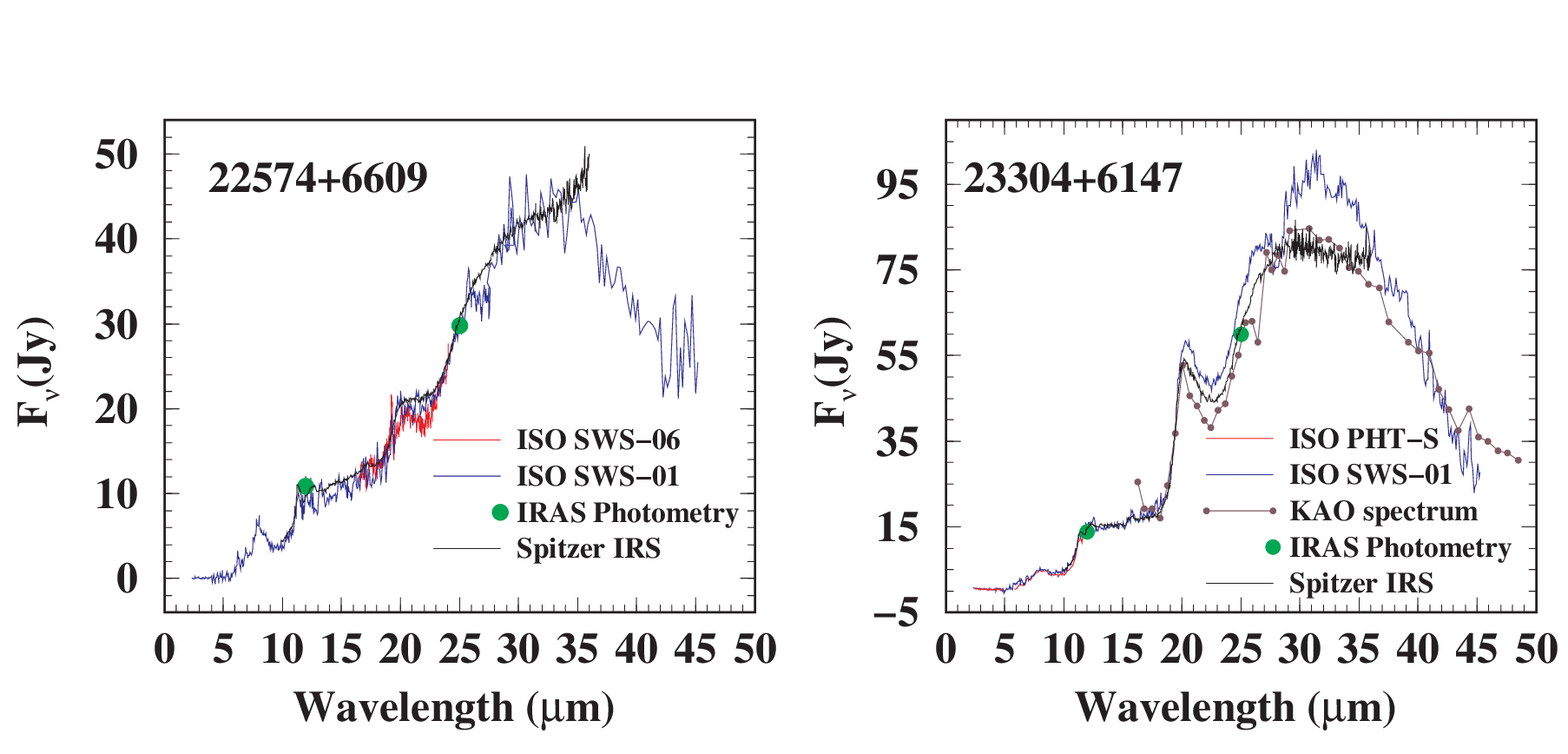}\figcaption[] {Comparison
of {\it Spitzer}, {\it ISO}, and KAO spectra of IRAS 22574+6609
and IRAS 23304+6147.
\label{fig3}}
\end{figure}

IRAS 23304+6147 is a well-documented 21 $\mu$m source, being  one
of the original four {\it IRAS} sources in which this feature was
detected \citep{kwo89}.  The 30 $\mu$m feature was discovered by
\citet{omo95} using the KAO.  Good observations were subsequently
carried out with {\it ISO} \citep{vol99,vol02}, and we observed it
again partly for comparison purposes.  Shown in Figure \ref{fig3}
are the {\it ISO} SWS01 (3$-$45 $\mu$m) and PHOT-S (3$-$8 $\mu$m)
and the KAO (16$-$48 $\mu$m) spectra 
along with the {\it Spitzer} spectrum. It can be seen that
the spectra are very similar in intensity from 10$-$19.5 $\mu$m and
similar in  shape from 10$-$27.5 $\mu$m. Beyond this, out to 36
$\mu$m, the {\it ISO} spectrum is about 10$\%$ higher.  
The KAO spectrum is roughly similar in flux to the {\it Spitzer} spectrum 
beyond 27 $\mu$m, but neither it nor the {\it ISO} spectrum shows the
flattening seen in the {\it Spitzer} spectrum beyond 33 $\mu$m.
The new {\it Spitzer} spectrum is noticeably flat between 30 and 35
$\mu$m.

\placefigure{fig3}

The {\it Spitzer} spectrum was convolved with the {\it IRAS} bands
to calculate synthetic {\it IRAS} photometry. This resulted in a
25 $\mu$m flux of 59.9 Jy, in excellent agreement with the {\it
IRAS} 25 $\mu$m flux of 59.1 $\pm$ 4$\%$ (53.7 Jy
color-corrected). The color-corrected {\it IRAS} 12 and 25 $\mu$m
flux measurements have also been included for comparison in Figure
\ref{fig3}.

{\it IRAS 05113+1347}: The spectra in each of the two nod
positions in SH and LH agreed with each other and were averaged.
To combine SH and LH smoothly required that we increase the SH
spectra by 7$\%$.  The spectrum shows a moderately strong 21
$\mu$m feature and a strong 30 $\mu$m feature.  At the shorter
wavelengths are seen emission features at 11.4 and 12.1 $\mu$m and
perhaps weak features at 13.3 and 15.8 $\mu$m, along with a weak 
feature at 22.3 $\mu$m.

The 21 $\mu$m feature was first detected in this object in a low S/N 
CGS3 spectrum from UKIRT following the suggestion of it in the {\it
IRAS} spectra \citep{kwo95}, but the source was not observed with
{\it ISO}. The {\it Spitzer} synthetic {\it IRAS} photometry
resulted in a 25 $\mu$m flux of 14.2 Jy, 7$\%$ lower than the {\it
IRAS} 25 $\mu$m flux of 15.3 $\pm$ 6$\%$ (12.6 Jy
color-corrected).

{\it IRAS 05341+0852}:  The SH and LH spectra in the two nod positions 
agreed reasonably well with each other and were averaged.  
The flux levels in most of the SH orders were scaled 
by a few percent to agree with the adjacent orders, but the data 
for the first three orders (10$-$12 $\mu$m) agreed well without any shifts.
The SH and LH spectra were then combined without any shift in flux.
The spectrum shows a strong 11.3 $\mu$m feature, a moderately strong 21
$\mu$m feature, and a strong 30 $\mu$m feature.  

The 21 $\mu$m feature in this source was also first detected in CGS3
spectra from UKIRT \citep{kwo95}; the source was not observed with
{\it ISO}. 
Present is a feature at 12.4 $\mu$m, and 
a broad but weak feature at $\sim$15.8 $\mu$m.
Also seen is a very narrow emission feature at 13.7 $\mu$m
that agrees with the position of the C$_2$H$_2$ molecular band seen in
absorption in carbon stars. 
The simulated {\it IRAS} photometry value was found to be 9.4 Jy at 
25 $\mu$m, compared to the observed  {\it IRAS} value of 
9.9 $\pm$ 7$\%$ Jy (8.7 Jy color-corrected).

{\it IRAS 06530$-$0213}: The spectrum was good at only one nod
position in each of SH and LH.  The longest wavelength order in LH
was scaled up by 8$\%$ to match the rest of the LH spectrum. The
SH spectrum was scaled up by 24$\%$ to merge the two spectra; this
is presumably to correct for flux missed due to poor centering of
the SH spectrum.
(Note that if poor centering is indeed the cause of the flux loss, 
the proper correction is wavelength dependent and not 
simply a constant scale factor.  However, we do not have enough 
information to make such a correction; thus this likely results 
in an uncertainty of several percent in the shape of the SH spectrum.)
The 21 $\mu$m feature is strong; the 30 $\mu$m feature is not
as relatively strong as in IRAS 23304+6147 and it shows an
unusual inflection around 33 $\mu$m. Also seen are AIB emission
features at 11.4 (strong), 12.3, and 13.2 $\mu$m, a strong,
medium-width feature at 15.8 $\mu$m, and a weak 
feature at 22.3 $\mu$m; the apparent feature at 14.6 $\mu$m
is likely an artifact.  There appears to be a weak, very narrow emission
feature at 13.7 $\mu$m, the position of the molecular C$_2$H$_2$
band.

This object had no previous mid-infrared spectrum, so this is the
discovery spectrum of the 21 and 30 $\,\mu$m emission features in
this object.   The {\it Spitzer} flux is noticeably lower than the
{\it IRAS} flux, with a synthetic 25 $\mu$m flux of 21 Jy as
compared with the {\it IRAS} flux of 27.4 $\pm$ 4$\%$ Jy (25.7 Jy
color-corrected). This large difference in flux is attributed to
flux loss in the {\it Spitzer} spectrum due to the non-optimal
centering of the object in the slit.

{\it IRAS 07430+1115}: In both the SH and LH spectra the two nod 
positions are in good agreement and were averaged.
The SH and LH spectra were combined with no shift in flux,
but the very shortest part of the LH spectra was removed due to lack of 
agreement with the SH in part of the region of overlap.
Seen is a 21 $\mu$m feature and a strong 30 $\mu$m feature.  
The 11.3 $\mu$m feature is seen a broad plateau extending out to 12 $\mu$m.
The {\it Spitzer} flux is close to the {\it IRAS} flux, with a 
synthetic 25 $\mu$m flux of 27.9 Jy, close to the {\it IRAS} flux 
of 29.9 $\pm$ 5$\%$ Jy (24.7 Jy color-corrected). 

The earlier UKIRT CGS3 spectrum did not reveal the 21 $ \mu$m feature \cite{hri99}, 
but since the CGS3 spectrum only extended to 24 $\mu$m 
and at lower S/N, it is not surprising that this weak feature was not detected.
This target was not observed with {\it ISO}.
Thus this is the discovery spectrum of both the 21 and 30 $\mu$m features.
Based on the earlier CGS3 spectrum, it had appeared that IRAS 07430+1115 
was an exception to the rule that all C-rich PPNs are 21 $\mu$m sources.
This exception is now removed.

{\it IRAS 19477+2401}:  Only one of the LH nod positions contained
a good spectrum.  Neither of the SH nods had a useable spectrum;
it was on the edge of the slit and appeared very faint.  These
problems were traced to poor positioning of the object in our
spectrum and mis-identification of the source in the SH archival
observation.  The originally published coordinates were not of
sufficient accuracy, but a correct finding chart was shown
\citep{hri85}.
In the one LH spectrum that we used, there was still a need to
scale some of the individual orders by a few percent before
combining them.

The resulting spectrum is shown in Figure 1.  It shows 21 $\mu$m
feature and a strong 30 $\mu$m  feature. In addition, there is a
weak feature emission at 22.3 $\mu$m and possible weak emission
features at 28.6, 29.8, 32.6, and perhaps even 34.9 $\mu$m.  At
31.4 $\mu$m there is a narrow dip followed by a narrow peak; since it appeared
in two separate orders, we did not remove it from the spectrum.
However, it would be quite unusual to see a P Cygni profile at this
spectral resolution.  The 30 $\mu$m  feature shows a slight inflection in the
spectrum around 33 $\mu$m.  The synthetic {\it IRAS} 25 $\mu$m
flux is 50.5 Jy, 8$\%$ less than the {\it IRAS} value of 54.9
$\pm$4$\%$ (51.2 Jy color-corrected).

IRAS 19477+2401 was observed with CGS3 but the 21 $\mu$m feature
was not seen in the noisy 20 $\mu$m spectrum \citep{kwo95}, and
the {\it ISO} spectrum shows flux problems also attributed to an
inaccurate position \citep{hri00}.  This {\it Spitzer} spectrum is
the discovery spectrum for the 21 $\mu$m  and the 30 $\mu$m
emission features in this source.

{\it IRAS 22574+6609}: In both SH and LH, the spectra in the two
nod positions agree well and were averaged.  In the LH, we shifted
the longest order (LH11) up by 6$\%$ to match with the previous
order, and the SH was scaled up by 2$\%$ to match with the LH. The
spectrum shows a moderately strong emission features at 21 $\mu$m
and 30 $\mu$m.  A very strong emission feature is seen at 11.3
$\mu$m, along with a medium-width feature peaking at 12.6 $\mu$m and
another one at 17 $\mu$m (about 0.5 $\mu$m wide).
The is evidence of a weak absorption feature at 13.7 $\mu$m that
may be the molecular band due to C$_2$H$_2$.

The 21 $\mu$m feature was first detected in this object in the
{\it IRAS} LRS spectrum \citep{hri91b}, followed by observation of
the 21 and 30 $\mu$m and AIB features with {\it ISO} SWS01
\citep{hri00} and SWS06 spectra \citep{vol99,vol02}.  These
previous spectra were rather noisy and the new {\it Spitzer}
spectra show the 21 and 11.3 $\mu$m features at much higher S/N.
In Figure \ref{fig3} are compared the {\it Spitzer} and {\it ISO}
spectra.  They have the same general shape out to 35 $\mu$m, but
the {\it Spitzer} spectrum is $\sim$20$\%$ higher shortward of 23
$\mu$m. The spectral features differ in the region 12 to 14
$\mu$m; the {\it Spitzer} spectrum shows a feature at 12.6 $\mu$m
while the noisier {\it ISO} spectrum shows features at 12.0 and
13.3 $\mu$m.  The {\it Spitzer} synthetic 25 $\mu$m flux is again
less than the {\it IRAS} flux, 27.3 Jy compared with 29.5
$\pm$4$\%$ Jy (29.8 Jy color-corrected).

We note that in most of the sources (IRAS 06530$-$0213, 07430+1115, 
19477+2401, 22574+6609, 23304+6147) the spectra are flat or even show 
an upturn in flux beyond $\sim$33 $\mu$m.  In the previous spectra 
of IRAS 23304+6147 and 22574+6609 this was not seen.  As mentioned 
in the discussion of the individual sources, the longest wavelength 
order, LH11 (34$-$36 $\mu$m), was scaled upward by 5$-$8 $\%$ 
of some cases to match the rest of the LH spectrum.  
Although the data are of high S/N, we maintain some skepticism about 
the accuracy of the shape of the spectra in the 33$-$36 $\mu$m region.  
Perhaps slit losses make some contribution the the suspected problem, 
since some of the sources are  know to be extended in the N band.

\section{DISCUSSION OF THE SPECTRAL FEATURES}
\label{results}

\subsection{21 $\mu$m Emission Feature}

\citet{vol99} used {\it ISO} spectra to examine the 21 $\mu$m
feature in eight PPNs, including IRAS 23304+6147 and 22574+6609.
They found the feature to have a peak wavelength of 20.1 $\pm$ 0.1
$\mu$m with a rather similar shape in all eight sources.  However,
they found a large range in the relative strength of the feature
(peak-to-continuum ratio).

\begin{figure}
\plotone{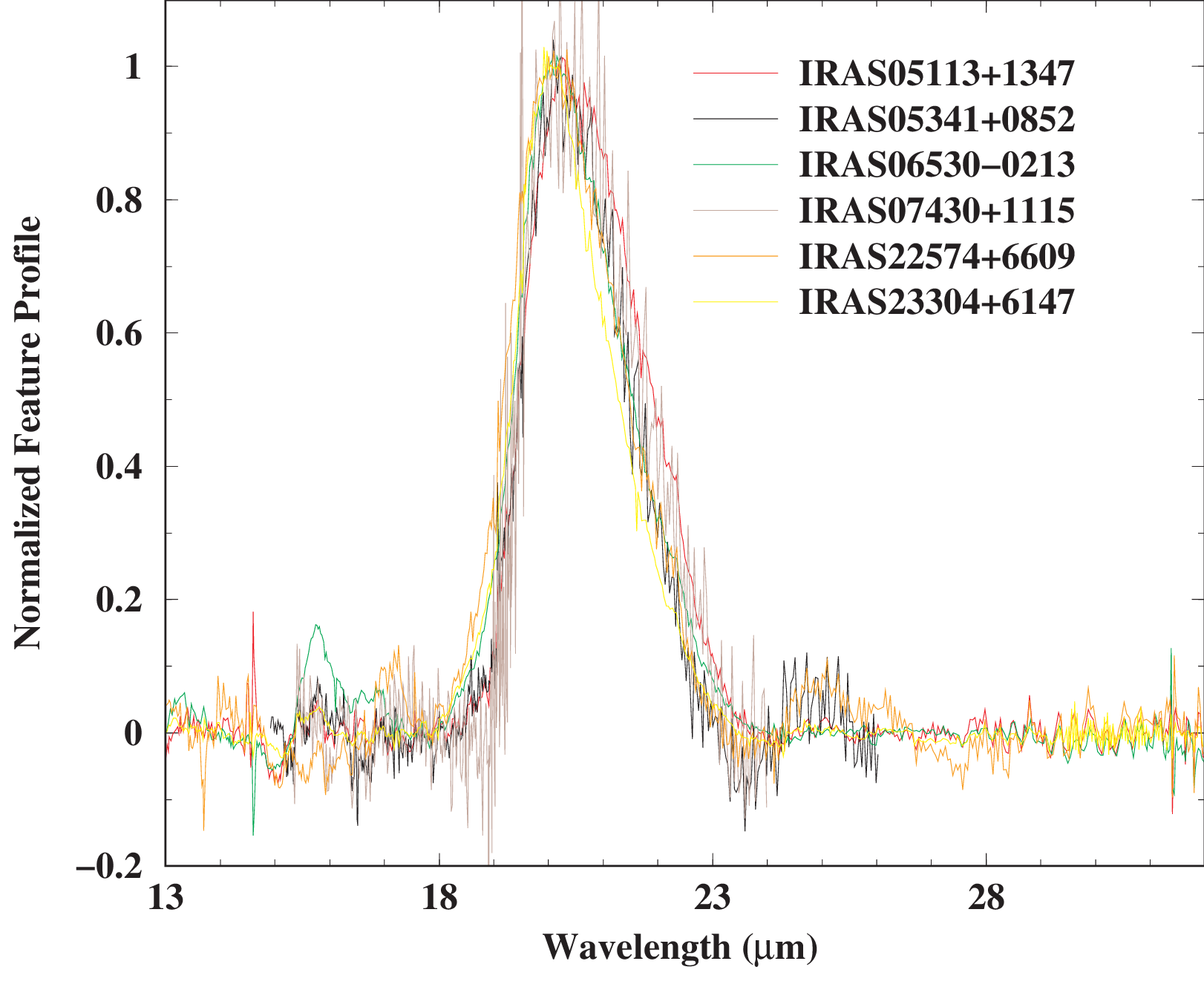} \figcaption[] {Comparison of the normalized 21
$\mu$m spectrum. (Excluded is IRAS 19477+2401 due to the absence 
of the spectrum shortward of 18 $\mu$m with which to set the continuum.) 
\label{fig4}}
\end{figure}

We can similarly compare the feature profile in these new, higher
S/N {\it Spitzer} spectra.  This was done for the six objects
with spectra in both SH and LH modes (for continuum fitting).
To determine the continuum, we used the flux at the shorter and longer 
wavelength sides and fit it with a spline function that gave a good 
smooth fit to this continuum; these ranged in order from 5 to 8.  
For three of the sources, IRAS 06530$-$0213, 22574+6609, and 
23304+6147, we treated as the continuum the flux at 13$-$18 $\mu$m 
and at 24$-$33 $\mu$m (which also includes the part of the 30 
$\mu$m feature), while for the other three, for which there was 
a more steeply rising continuum on the long-wavelength side, 
we used a smaller range about the feature. This function was then 
divided into the spectrum to produce the feature profile. 
The relative strength of the feature varied quite
a bit, from 0.21 to 1.46. An approximately similar range was
determined in previous studies of the brighter objects
 \citep{vol99, hri00, vol02}; however these present spectra are
superior due to their higher S/N. These
peak-to-continuum ratios are listed in Table \ref{21pro}, along
with the feature width and peak wavelength for each source. The
feature profiles are similar; this is shown graphically in Figure
\ref{fig4}, where the profiles have been normalized to 1.000 and
the continuum removed.
The long wavelength side of the feature differs slightly among the sources,
especially for IRAS 05113+1347.  However, this depends sensitively
on the 30 $\mu$m feature, and the assumption that there is a region of 
continuum between the two features and thus that the 30 $\mu$m 
feature does not contribute flux between 21 and 24 $\mu$m.  
The short wavelength side of the 21 $\mu$m feature is less susceptible 
to contamination by strong features and appears to be relatively 
consistent among the sources.
The peak wavelength of the feature is 20.1$\pm$0.1 $\mu$m,
the same as found for the sources in the earlier {\it ISO} study.  The feature
width we find is greater than in the earlier study, but this is
likely due to the fact that we had more continuum to work with.
IRAS 23304+6147 was included in both studies, and its 21 $\mu$m 
feature is wider by 20$\%$ in the {\it Spitzer} spectrum.

\placetable{21pro}

\placefigure{fig4}

In all of the 21 $\mu$m sources studied in detail, this feature appears to 
have a relatively similar central wavelength and profile shape in each, 
but different strengths.  Uncertainties arise in 
defining the feature since it sits upon a continuum and adjacent to 
a very broad 30 $\mu$m feature, both of which are rising steeply 
toward longer wavelengths. 

All carbon-rich PPNs that have been observed possess 
the 21 $\mu$m emission features \citep{hri08}.  
The sole exception to this statement may be IRAS 01005+7910 (OBe).
The low S/N {\it ISO} spectrum of this source may or may not possess the 
21 $\mu$m feature; it does possess the 30 $\mu$m and AIB features 
\citep{hri00}.  
Since this is the hottest know carbon-rich PPN and since the feature 
appears to be absent or weak in PNs, it may be that IRAS 01005+7910 
is displaying the expected weakening of this feature with evolution toward 
a hotter central star. 
The next hottest carbon-rich PPN star is IRAS 16594$-$4656 (B7e), and
it clearly shows the feature \citep{garlar99}.

The carrier of this feature is still not identified.  There have
been many suggestions, including PAHs, hydrogenated fullerenes,
nanodiamonds, and more recently TiC nanoclusters \citep{von00},
doped SiC \citep{spe04,jia05}, and the interaction of Ti atoms
with fullerenes \citep{kim05}.  \citep[For a review of these and other
potential carriers and additional references, see][]{pos04}.
Some of these predict features at other wavelengths, but none of
them have been confirmed.  SiC does indeed produce an observed
feature at 11.3 $\mu$m, but it does not have the correct flux
ratio with the 21 $\mu$m feature \citep{jia05}.  We note below a
possibly correlated feature at 15.8 $\mu$m in our observed
spectra.

\subsection{30 $\mu$m Emission Feature: Two Components?}

We have previously used {\it ISO} spectra to study the 30 $\mu$m
feature in PPNs \citep{hri00,vol02}.  In these {\it ISO} spectra it appeared
that this very broad feature was resolved into two separate
features at 26 and 33 $\mu$m, seen with varying strengths in PPNs.
These new {\it Spitzer} spectra do not support this claim. This
can be seen by the comparison of the {\it Spitzer} spectrum with
the {\it ISO} spectrum for IRAS 23304+6147 
as shown in Figure \ref{fig3}a. The apparent downward inflection in the
{\it ISO} spectrum at 27.5$-$29.5 $\mu$m is evidently due
completely to the unreliable flux of {\it ISO} band 3E
(27.5$-$29.5 $\mu$m).  Also, the data in {\it ISO} band 4
(29.5$-$45 $\mu$m) are known to be very noisy due to the many
cosmic ray hits and the flux calibration is less reliable. 
A similar effect is seen in the comparison of the Spitzer spectrum 
of IRAS 22574+6609 with its {\it ISO} spectrum 
as shown in Figure \ref{fig3}b. 
The downward inflection in the {\it ISO} spectrum
between 26 and 33 $\mu$m is not seen in the {\it Spitzer} spectrum
of IRAS 22574+6609, nor is it seen in any of our other {\it
Spitzer} spectra of sources with the 30 $\mu$m feature.
Low-resolution {\it Spitzer} spectra of evolved carbon stars in
the LMC also do not show an indication of the 30 $\mu$m feature
being resolved \citep{buc06}, although this feature does appear to
be significantly weaker in these LMC sources than in similar
objects in the Milky Way Galaxy or in these PPNs. 
\citet{hon02} used {\it ISO} spectra to study the 30 $\mu$m feature 
in a large sample (63) of carbon-rich evolved stars $-$ AGB, post-AGB, 
and PN.  They treated it as a single feature and found that a good fit to 
the 30 $\mu$m profile could be obtained using MgS grains with a 
temperature differing from the continuum temperature.  However, 
in $\sim$40\% of these a residual excess existed at 26 $\mu$m.  
Whether this excess is also an artifact due to the above problems 
in the {\it ISO} spectra might be further investigated using 
{\it Spitzer} spectra of these {\it ISO} targets.
In a recent paper, \citet{zhu08} present a strong case for the feature 
to be due to MgS as a coating on SiC grains.
They go on to show how the feature shape and peak would change 
with the relative thickness of the coating, with the main peak 
at 25$-$28 $\mu$m and predicting a 
sub-peak at 33$-$38 $\mu$m when the coating is not too thick.

\subsection{AIB Emission Features}

A strong 11.3 $\mu$m feature, along with several corresponding 
features in the 12$-$14 $\mu$m region and an underlying plateau.
The known features at 11.3, 12.1, 12.4, and 13.3 $\mu$m are due to out-of-plane
bending modes of aromatic compounds with peripheral H atoms
attached to the aromatic rings. 
Although the carriers of these features are frequently attributed to 
PAH molecules, the strong associated underlying broad 
continuum suggests that the carrier of the continuum emission cannot 
arise from small, gas-phase molecules, and has to be emitted by 
solid-state grains or very large molecules with thousands of C atoms.

\begin{figure}
\plotone{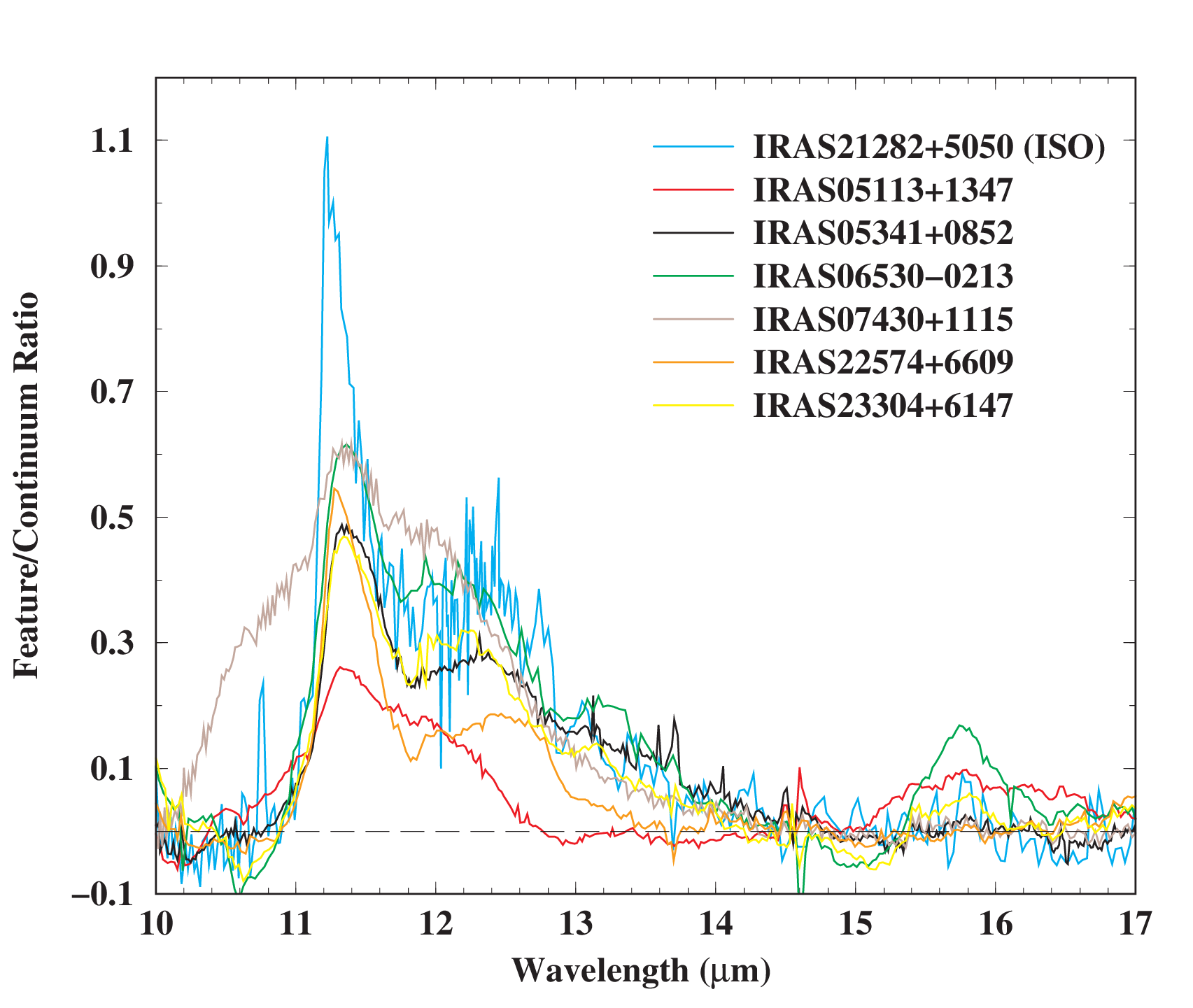} \figcaption[] {Comparison of the normalized
10$-$17 $\mu$m spectrum for the sources in this study.
Included for comparison is the {\it ISO} spectrum of the young PN
IRAS 21282+5050.  Seen in all the spectra is the strong emission
feature at 11.3 $\mu$m; emission at 12.3 $\mu$m is seen in most
and 13.3 and 15.8 $\mu$m in some of them.  (The spectra are noisy
in the range of 14.5 to 14.6 $\mu$m.) \label{fig5}}
\end{figure}

To investigate the shapes of these features, we show in Figure \ref{fig5} 
the 10$-$17 $\mu$m spectral region with the continuum removed 
(by division of a fourth-order cubic spline fitted from 
10.0$-$10.5 $\mu$m to 14$-$18 $\mu$m). 
The strong 11.3 $\mu$m feature has a relatively consistent
peak wavelength of 11.33 $\pm$ 0.03 $\mu$m (ranges from 11.27 to 11.36 $\mu$m
in these six PPNs), 
Shown for comparison is the {\it ISO} spectrum of IRAS 21282+5050, 
a young PN with a [WC] central star.
Its spectrum in this region has a general similarity to that of the PPNs. 
However, the 11.3 $\mu$m feature is narrower and peaks at a 
shorter wavelength, 11.21 $\mu$m.  
The peak of the 11.3 $\mu$m feature at shorter wavelengths is what is also seen in 
most other sources.
From a sample of 15 objects, consisting of HII regions, young stellar objects, 
and PNs, \citet{hon01a} found that they all had a profile peak very 
near to 11.23 $\mu$m.
\citet{vandie04} defined a classification scheme for the 11.3 $\mu$m feature that 
depended upon the shape and peak wavelength of the feature, but all three of their 
classes (A, AB, B) had peak wavelengths ranging from 11.20 to 11.25 $\mu$m.
However, \citet{slo07} found in a sample of objects with cooler central stars that 
the central wavelengths are shifted to the red relative to the above peaks found for
the class A and B sources.  They include class C AIB features, based on the classification of 
\citet{pee02}, and these peak at wavelengths similar to what we find.
\citet{slo07} attribute the difference in the class C AIB spectra as due to the 
fact that ``the carbonaceous material has not been subject to a strong ultraviolet 
radiation field, thus allowing the relatively fragile aliphatic materials to survive.''
Our targets all have cool central stars and aliphatic features (see below), and thus
are consistent with this suggestion. 


\placefigure{fig5}

Around 12 $\mu$m, a broader emission feature is seen in all of these spectra, 
with a peak wavelength ranging from 12.0 to 12.6 $\mu$m.  
In IRAS 06530$-$0213 and 23304+6147 and perhaps 05113+1347, a weaker 
emission feature is seen at 13.2$-$13.3 $\mu$m.  
The features at 12.1, 12.4, and 13.3 $\mu$m have been identified as the 
duo, trio, and quatro out-of-plane bending modes of C$-$H bonds 
and are associated with the 11.3 $\mu$m solo-CH feature \citep{hud99}.
In previous studies of the infrared spectral characteristics of
PPNs, we found that unlike PNs and HII regions, PPNs show
aliphatic emission features at 3.4 $\mu$m \citep{hri07} and 6.9, 
in addition to the features at 12.1, 12.4 and 13.3 $\mu$m which 
are uncommon in PNs or in other galactic nebulae. 
This spectral difference is suggested to be the
result of chemical evolution, with a progressive formation of
larger and larger clusters of aromatic rings and the removal of
peripheral aliphatic groups \citep{kwo99b}.
Absent in these PPNs is the relatively strong, asymmetric 12.7 $\mu$m feature found in many 
objects with AIB features, including IRAS 21282+5050 \citep{hon01a, slo05, slo07}.
In this sample, it is found only in IRAS 06530$-$0213.

The existence of broad plateau emission features at 8 and 12
$\mu$m are also a common feature among the PPNs but not
PNs and objects with hot central stars  \citep{hon01}.  This emission
plateau is attributed to a collection of out-of-plane bending
modes of a miscellaneous mixture of aliphatic side groups attached
to the aromatic rings \citep{kwo01}.  From Figure \ref{fig5}, we
can see that the 12 $\mu$m plateau feature is present among most
of the PPNs observed here, with IRAS 05113+1347 perhaps being an exception.  
The width of this feature is estimated to be about 2 $\mu$m. 
An exception here is IRAS 07430+1115, which appears to have 
a broader width extending to 10 $\mu$m, although it is hard to 
set the continuum at the short end of this spectrum.

\subsection{15.8 $\mu$m Emission Feature}

A relatively strong and medium-wide ($\sim$1.3 $\mu$m) feature is seen at
15.8 $\mu$m in two of the sources, IRAS 23304+6147 and
06530$-$0213. The feature also appears to
be present but weak in IRAS 05113+1347 and 05341+0852. 
\citet{mou99} point out features at 15.8  and 16.4 $\mu$m in their spectrum 
of NGC 7023, noting that their widths of $\sim$0.15 $\mu$m are similar to that 
of the 11.3 $\mu$m feature.
In the more extensive discussion of sources with features between 15 and 20 $\mu$m 
by \citet{vanker00}, this feature can be seen in two or three sources but is not 
particularly distinct and is usually weaker than the apparently associated 
feature at 16.4 $\mu$m, which is not present in our spectra.  These and the 
complex of other features which are thought to make up a 15$-$20 $\mu$m 
emission plateau in carbon-rich sources are attributed to C-C-C  
bending modes in PAHs.  
\citet{sel07} point out additional emission features in NGC 7023 at 17.4 and 
17.8 $\mu$m, along with a broad plateau feature at 17 $\mu$m and a weaker 
features at 18.9 $\mu$m.  Our C-rich PPNs show 3.3, 6.2, 11.3 $\mu$m 
features similar to those seen in NGC 7023 but none of these other, new features.
The 15.8 $\mu$m features seen in their spectra are not nearly as broad 
or strong as the 15.8 $\mu$m features seen in several of our new PPN spectra,
and thus they are not the same.

The width of this feature suggests that it is made up of a larger molecule 
than that producing the AIB features and perhaps is due to a solid-state 
feature.
To quantify the properties of this feature, we determined it peak wavelength and 
the relative strength of the feature (peak-to-continuum ratio).  For the weaker
sources, we determined only an upper limit to the relative strength.
The values are as follows: IRAS 06530$-$0213 (15.75 $\mu$m, 0.22), 
23304+6147 (15.79 $\mu$m, 0.12), 05341+0852 (15.77 $\mu$m, 0.06), 
05115+1347 (..., $<\sim$0.03), 22574+6609 (..., $<\sim$0.03), and 
07430+1115 (..., $<$0.02).
In the source with the strongest feature, IRAS 06530$-$0213, it has a width 
of 1.3 $\mu$m (15.1 to 16.4 $\mu$m).  The shape in IRAS 23304+6147 
is consistent with this and the shape in the weaker sources is also consistent, 
but the feature in these is of much lower S/N.

As can be seen by comparison with Table \ref{21pro}, the two sources with the 
much stronger 15.8 $\mu$m feature are also the sources in our study 
with the much stronger 21 $\mu$m feature. 
This raises the possibility of a possible relationship between these two features.  
An examination of the lower S/N {\it ISO} spectra of other 21 $\mu$m
sources shows that the 15.8 $\mu$m feature is likely present in IRAS
22272+5435, 19500$-$1709, 16594$-$4656, 23304+6147 
(as confirmed in this {\it Spitzer} spectrum), and in the low S/N 
spectrum of IRAS 07134+1005, the source with the relatively 
brightest 21 $\mu$m feature \citep{vol02,hri00}.
This possible relationship is based on only a few objects 
and so is only tentative, but it would certainly be worthwhile to investigate 
it for more of the 21 $\mu$m sources.  The presence of a feature correlated 
with the strength of the 21 $\mu$m feature would provide an important clue 
to help identify the carrier and could at minimum help to constrain the 
viable suggestions.

\subsection{13.7 $\mu$m C$_2$H$_2$ Feature}

We have previously mentioned a narrow (unresolved) feature at 13.7 $\mu$m
in three of these PPNs, in absorption in one (IRAS 22574+6609) and in 
emission in two (IRAS 05341+0852 and 06530$-$0213). 
This is at the wavelength of the molecular C$_2$H$_2$ (acetylene)
$\nu$5 bandhead feature 
commonly seen in absorption in carbon stars \citep{wil88,cer99,vol00}.
This feature may also be present in IRAS 23304+6147 in both absorption and
emission (P Cygni profile; see Figure \ref{fig5}) and in IRAS 07430+1115 in 
emission. This C$_2$H$_2$
feature has previously been reported in absorption in only three
galactic PPN, AFGL 618 \citep{cer01}, AFGL 2688, and IRAS 13416$-$6243 
\citep{kra06}; it has recently been  also seen in absorption in the LMC PPN/PN 
object SMP LMC 11 \citep{ber06} and the SMC PPN object MSX SMC 029 
\citep{kra06}. 
\citet{kra06} and \citet{slo07} point out that when the C$_2$H$_2$ 
absorption feature appears in sources with PAH emission, the sources are those 
classified as the rare class C PAH sources \citep{pee02}.
Our sources with C$_2$H$_2$ would all be classified as class C PAH sources  
based on their broad 8 $\mu$m features 
and secondarily based on the longer wavelength of their 11.3 $\mu$m feature, 
and thus this claimed correlation is supported by our new observations.

To our knowledge, C$_2$H$_2$ has previously been seen in emission in only one
evolved stellar object, the AGB star IRC+10216 \citep{kea93}. 
Thus this is the first report of the molecule in emission in a post-AGB object.
One might consider the emission to be due to collisional excitation.  However, 
collisionally-excited H$_2$
emission is not seen in any of these three PPNs at 2.12 $\mu$m
(1$-$0 S(1)) \citep{kel05}, nor is H$_2$ emission seen in any
objects in this study at 12.27 $\mu$m (0$-$0 S(2)) or 17.03 $\mu$m
(0$-$0 S(1)). (Only for IRAS 22574+6609, with a 13.7 $\mu$m absorption feature, 
is a (weak) 2.12 $\mu$m H$_2$ emission feature seen.)  
We have recently confirmed the presence of C$_2$H$_2$ $\nu$5 lines having 
P-Cygni profiles in three of these objects using high resolution spectroscopy, as 
will be discussed in more detail elsewhere.
This feature has the potential to serve as a diagnostic of the physical conditions 
in the circumstellar envelope.


\subsection{Summary of Spectral Features}

We summarize the presence of these spectral features in our
targets in Table \ref{features}.  Also included are the
carbon-to-oxygen (C/O) abundance ratio and observations of AIB
features at shorter wavelengths as found in other studies.  It can
be seen that all of the sources in this sample with the 21 $\mu$m
feature are carbon-rich, possess the 30 $\mu$m emission feature,
and also possess AIB emission features in the 3$-$14 $\mu$m range
(except IRAS 19477+2401, which has not been observed in high S/N
spectra in this range).  These spectral features appear to accompany
the presence of the 21 $\mu$m feature.  All of the sources in this study 
that have visible spectral types also possess C$_2$ and in most cases
C$_3$ in absorption.
In addition to the in-some-cases-strong emission feature at 15.8 $\mu$m, 
there is an unidentified weak feature seen at 22.3 $\mu$m in four of these sources.

\placetable{features}

\section{MODELING OF THE SEDS}
\label{sed_models}

To learn more about the physical conditions under which these features arise 
(T, $\rho$), and also to determine the mass loss rates ($\dot M$) and 
expansion time scales (t$_{\rm dyn}$), 
we carried out modeling of the circumstellar envelopes.
Model fits to the spectral energy distributions (SEDs) were
carried out for the six sources with complete 10$-$36
$\mu$m spectra.
The continuum was assumed to be due to amorphous
carbon (AC) grains, with optical constants by \citet[type
AC2]{rou91} and an assumed size of 0.1 $\mu$m. 
The radiation transfer code by \citet{leu76}, DUSTCD, was used to
produce one-dimensional models. The dust envelope is assumed to
have an inner radius r$_{\rm in}$, a dust density profile of
1/r$^{\alpha}$, and to extend to 0.75 pc 
(somewhat arbitrarily set at the distance reached in 5 x 10$^{4}$ yrs 
with an expansion rate of 15 km s$^{-1}$).  
 The temperature of the central star (T$_*$) was initially set at the
value derived in published atmospheric model analyses and then adjusted
slightly ($\le$250 K) to fit the SED at shorter wavelengths.  A Kurucz
model atmosphere \citep{kur93} with log~{\it g} = 1.0
was used for the central star, and this was the
assumed energy source for the nebula.  The interstellar extinction A$_{\rm V}$
was treated as an adjustable parameter.  
The other parameters that were initially adjusted were the temperature (T$_{\rm d}$) 
and inner radius of the cool dust shell, the optical depth ($\tau$$_{\rm o}$), 
and the exponent of the density law ({$\alpha$}).
These were adjusted by trial-and-error fits of the calculated to the observed SEDs  
until a ``best fit'' was determined.  We began by fitting the mid-infrared
continuum at 18 $\mu$m and 100 and/or 60 $\mu$m; we set T$_{\rm d}$ 
to the highest value that did not go above the continuum and adjusted 
r$_{\rm in}$ and {$\alpha$}.
We then adjusted the optical depth in the near-infrared, along with A$_{\rm V}$ 
and T$_*$, to fit the near-infrared and also the visible as well as possible.
These parameters are not strongly coupled, with each primarily effecting a distinct 
aspect of the SED. 
More details of the basic model are given by \citet{hri00}.

After initial attempts with this basic model to fit the SEDs of
these PPNs, 
the model was adjusted in two ways.  
First, in some cases it was found that a single dust shell with 
the high density needed to fit the mid-infrared SED and a large radial 
extent produced flux that was too high to fit the long wavelength
60 and 100 $\mu$m {\it IRAS} fluxes, even if we assumed a very
steep density law ($\alpha$$\ge$4). Thus we added to the model 
a discreet thin, dense, cool (c) dust shell of inner and
outer radius r$_{in}$(c) and r$_{out}$(c), respectively.  
Second, it was found that for most of the sources, the model produced
flux that was too low to fit the observations in the M and L
bands.  To improve this fit, a component of hot dust was added
where needed.  This hot (h) component was of low density and
mass, and served to bring the model into better agreement with the
observed flux in the 3$-$5 $\mu$m region.  The hot dust component
begins close to the star, with a small value for r$_{in}$(h).
The values for T$_d$(h) and r$_{\rm in}$(h) were adjusted to fit the 
M and L observations.
The density laws for both the cool and hot components are assumed to
have the same exponent. The resulting density ratio of the cool to the hot
component ($\rho$(c)/$\rho$(h)) ranges among the models from 7700
to 15,000 at a radius of r$_{\rm in}$(c), and the hot dust contributes only 
1.5$-$11~$\%$ of the optical depth at 11.2 $\mu$m.

The AIB plateau features were not included in the modeling.
They are likely produced by some combination of flourescence and non-equilibrium 
heating, which the dust radiative transfer code does not have the capability 
to include.
However, they do not carry a large part of the total energy emitted in the 
mid-infrared and presumably therefore would not have much of an effect 
on the optical depth in the dust shell or the derived parameters. 
IRAS 22574+6609 is a possible exception to this generalization 
since it has an unusually high optical depth dust shell.

Empirical fits to the 21 and 30 $\mu$m
features were added to the continuum.
The 21 $\mu$m feature template is the same one we used previously 
to fit {\it ISO} spectra, derived from the {\it ISO} 
high-resolution SWS06 spectra of carbon-rich PPNs.
The 30 $\mu$m feature template is derived from the fitting of the 
IRS spectrum for IRAS 05113+1347, since this source has a strong and
relatively well-defined 30 $\mu$m feature.  
The long wavelength end of the feature was defined by a linear 
extrapolation of the observed feature shape from 32 to 36 $\mu$m, which goes to 
zero at 43 $\mu$m; that is reasonably consistent with the {\it ISO} long 
wavelength limit of the feature in PPN \citep{vol02,hon02}.  
(The extent to which this template fits or does not fit for the other 
objects is either telling us something about problems in the IRS data 
at long wavelengths or is indicating that there are variations in the 
feature shape.)  
We did not attempt to empirically fit the AIB plateau features.

\begin{figure}
\plotone{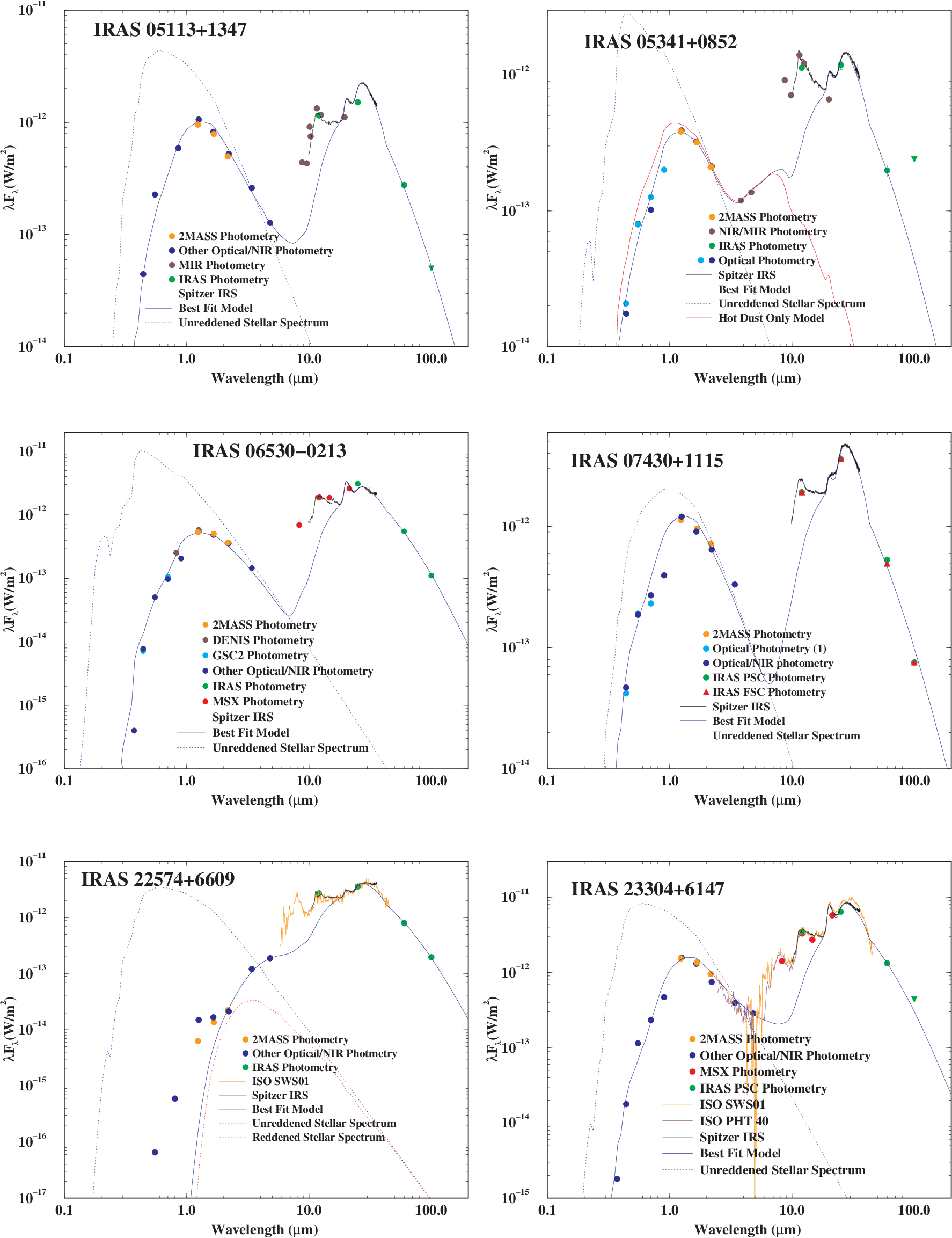}\figcaption[] {Fits of the models
to the SEDs.  Included are the ground-based (2MASS, DENIS, NIR=near-IR, MIR=mid-IR, Other),  
{\it MSX}, {\it IRAS}, and {\it ISO} data, in addition to the {\it Spitzer} IRS spectra. \label{fig6}}
\end{figure}

These model results are listed in Table \ref{models} and the fits
are shown in Figure \ref{fig6}.  In calculating the mass loss rate, 
a gas-to-dust density ratio of 330 was assumed.
Note that while the total mass of the enhanced shell is well determined,
the mass loss rate is less so, since a similar model fit could probably be 
obtained using a narrower or wider shell of the same mass.
Below we discuss briefly the model for each object separately.

\placetable{models}

\placefigure{fig6}

{\it IRAS 05113+1347.} $-$ The fit to the SED of this object
required a discrete cool shell.  Initial attempts to fit it with an
extended dust shell required a very high density exponent of
$\alpha$ = 5, and even then the fit was not as good as with a
discrete shell.  The hot dust component is needed to fit the
M-band flux.  This hot component is of low mass, and contributes
only 3$\%$ of the optical depth at 11.2 $\mu$m.  The model
provides a good fit to the SED, except for the emission plateau
from 8 to 14 $\mu$m and the 15.8 $\mu$m emission feature which 
were not included in the modelling.  
The visible image shows only the star, suggests that we may be looking
pole-on to the nebula \citep{uet00}.

{\it IRAS 05341+0852.} $-$ The SED of this object was fitted by
an extended, detached cool dust shell with a high density exponent of
$\alpha$ = 3.5.  To fit the L and M flux, a hot dust component is 
also needed.  This hot component contributes 11$\%$ of the optical depth 
at 11.2 $\mu$m but less than 0.03$\%$ of the total mass in the envelope.
The model fit to the SED is good, except for the emission plateau
from 8 to 14 $\mu$m and some excess emission observed from 30 to 36 $\mu$m.
The hot dust component leads to a particularly good fit to the L and M flux.
The visible image \citep{uet00} shows an elongated elliptical nebula around 
the central star and suggests that one is looking at a bipolar nebula at some 
intermediate orientation.

{\it IRAS 06530$-$0213.} $-$ The SED of this object can be fitted
by a normal, detached and extended cool dust shell with no extra
hot dust component.  The emission plateau from 8 to 14 $\mu$m and
the strong 15.8 $\mu$m emission feature were not fitted.  The
visible image \citep{uet00} suggests that one is looking at a
bipolar nebula at some intermediate orientation.

{\it IRAS 07430+1115.} $-$ The modeling of the SED of this object
was more difficult than most.  An adjustment of the density 
exponent to its best value resulted in $\alpha$ = 1, implying 
a decreasing mass loss rate.  This is both unlikely and contrary 
to the results for the others, so we chose to fix $\alpha$ = 2, 
implying a constant mass loss rate.  The resulting model fit was
only slightly inferior to that with $\alpha$ = 1.
The SED was fitted with a low density envelope and a high 
density cool shell region.  The observational excess in the 
L band suggests the presence of a hot component for this 
object also, but without an M-band observation, it is too 
unconstrained to model.
The model fit is not as good as most of the other sources.  
The visible and near-infrared photometry cannot be fit by a normal model 
atmosphere with reddening; the resulting compromise fit includes
more uncertainty in A$_V$ than in most of the models.
The model fit falls below the observations at 60 $\mu$m,
is not good on the long wavelength side of the 21 $\mu$m features, 
and falls below the L-band flux, as mentioned above.  Again, some excess emission 
is observed from 30 to 36 $\mu$m and the emission plateau from 8 to 14 
$\mu$m is not included in the model fit.
The visible image \citep{uet00} shows an approximately circular 
nebula around the central star.

{\it IRAS 22574+6609.} $-$ This source is very faint in the visible (V=21.3) and
it lacks a visible spectrum, so we assumed a typical PPN stellar
temperature of 5500 K.  The model requires a much higher optical
depth ($\tau_o$) for this object than the others, by a factor of 10$-$100.
The fit of the model to the SED of this
object is good from 18 to 28 $\mu$m and then from 60 to 100
$\mu$m. The emission plateau rises above the continuum from 8 to
18 $\mu$m. Beyond 28 $\mu$m, the IRS spectrum continues to rise.
An attempt to fit the IRS spectrum to 34 $\mu$m leads to a very poor 
fit to the 60 and 100 $\mu$m data.  Without spectra extending to 45
$\mu$m, it is hard to know exactly where the continuum lies. To
fit the K, L, and M band data, a hot component was added to the
model, which contributes less than 2$\%$ of the optical depth at
11.2 $\mu$m.  While this brings these data into agreement, we are
still left with poor agreement at the shorter wavelengths,
with extra flux observed below 2.2 $\mu$m and through
the visible.\footnote{The near-infrared measurements of IRAS
22574+6609 by \citet{hri91b} made with a 15$\arcsec$ aperture
likely include the nearby (5$\arcsec$ south) source 
2MASS 22591874+6625430, which is brighter in J but fainter in H and K.} 
Such behavior in dust shell models has been seen previously in one-dimensional 
models of extreme carbon stars including IRC+10216 
and AFGL 3068, and it is attributed to a non-spherical dust shell geometry.
This may also be the cause in IRAS 22574+6609.
The visible images show a faint, edge-on bipolar nebula with a dark
dust lane \citep{uet00,su01}.  
We suggest that the extra short wavelength emission is due a low
optical depth path (hole) out of the dust shell through which some
scattered light is escaping, similar to the situation in IRC+10216 
\citep{men02}. This may be consistent with the off-center intensity peak seen by 
\citet{san06} in their high resolution near-IR images of IRAS 22574+6609.
They show that this peak is not at the center of the nebula and thus is not
the central star seen directly.  
We have not tried to add extra complexity to the model simply to fit the 
visible wavelength portion of the SED.
These new model parameters are significantly
different from our previous model for this source.  However, those were based on 
much poorer quality {\it ISO} spectra, in which we chose to disregard the observed 
{\it ISO} spectrum from 27 to 45 $\mu$m \citep{hri00}.  
The new model was best fit with A$_{\rm V}$ = 0.0; 
however, galactic extinction studies by \citet{nic80} suggest a value 
$\ge$2.5, so we fixed A$_{\rm V}$ = 3.0.  The fit was slightly worse, 
but at a level that is negligible on Figure \ref{fig6}.  

{\it IRAS 23304+6147.} $-$ The model gives a good fit to the SED
apart from the emission plateau shortward of 18 $\mu$m.  A hot dust
component is required to fit the M-band excess.  This hot dust
contributes less than 2$\%$ of the optical depth at 11.2 $\mu$m.
The visible image \citep{sah07} is similar to that of IRAS
06530$-$0213, and also suggests a bipolar nebula at some
intermediate orientation.  This new model for IRAS 23304+6147 differs 
slightly from the one that we previously published based in {\it ISO} 
spectra \citep{vol02}, due primarily to the adoption of a stellar model rather 
than a black body for the central star.  

Thus it can be seen that the model fits are generally good (excluding the 
unmodelled AIB emission plateau from 8 to 14 $\mu$m), except for 
IRAS 22574+6609 and perhaps IRAS 07430+1115.  
Observations in the 5$-$10 $\mu$m range are needed 
to better constrain the continuum on the short wavelength side of 
the AIB emission plateau.
Although most of these PPNs are known to have a bipolar morphology, 
the assumption of spherical symmetry will have little effect on the fitting 
of the mid-infrared emission and the calculated mass-loss rate.  This is the 
case because the dust shells are optically thin in the infrared where the 
dust is detected.  
The only exception to this might be IRAS 22574+6609, where the 
derived  optical depth is much higher than for the other objects.  
In this case, the morphology might affect the mid-infrared emission and the 
resulting observed flux would then depend upon observing orientation.  

The cool dust temperatures are similar in these models, with values in the range 
120$-$150 K at the inner radius of the cool dust shell.  
Including an additional six 21 $\mu$m sources modelled earlier based on {\it ISO} 
spectra, the full range of T$_{\rm d}$(c) is 140 to 210 K 
\citep[disregarding the earlier model of IRAS 22574+6609;][]{hri00}.
The absence of C-rich PPNs with higher values of T$_{\rm d}$(c) arises partly from 
the absence of cooler, spectral type K central stars with smaller detached dust shells, 
even though there is no bias in detecting these \citep{vol89}.  
The absence of lower values of T$_{\rm d}$(c) at the inner radius of the 21 $\mu$m 
sources suggests that either the carrier is not excited at these lower dust temperatures 
or that it is destroyed due to the presence of the harder UV radiation field 
found when the central star has evolved to high temperatures. 
The value of T$_{\rm d}$(c) observed in a particular source depends 
sensitively on the rate of evolution of the central star as well as the expansion 
velocity of the detached shell. 
The temperatures for the hot dust are in the range $\sim$500$-$800 K. 
For IRAS 05113+1347, 05341+0852, and 23304+6147, 
the density of the cool dust is 10,000$-$15,000 times that of the hot dust,
but for IRAS 22574+6609, 
the ratio is only $\sim$200, due to the relatively larger amount of hot dust.

Many of the values in Table \ref{models} are scaled to the distance (D), and the
mass loss rate ($\dot M$) is also scaled to the expansion velocity (V). 
Assuming a luminosity of 8300 L$_{\sun}$, appropriate for a core mass of 0.63 M$_{\sun}$
\citep{blo95a} and published expansion velocities, the distances and mass loss rates 
for each object were calculated.  These are listed in  Table \ref{model_results}.
The mass loss rates range from a 10$^{-4}$  M$_\odot~{\rm yr}^{-1}$ to as high 
as a 6 $\times$ 10$^{-3}$  M$_\odot~{\rm yr}^{-1}$.  
These later values are very high and 
cannot be sustained very long in medium and low mass stars.  The total mass of the 
cool dust shells ($\Delta$M) range from 0.22 to 0.95 M$_\odot$.
The total mass of the enhanced shell is a more robust value than the mass loss rate since the width of the shell is not tightly constrained.  
The duration of the intense mass loss was calculated based on the shell radii and the 
expansion velocity, and ranges from 160 to 820 years;
again, these values, based on the width of the shell, are not tightly constrained.  
Thus the results of the model fits to the SEDs imply that these post-AGB objects 
lose most of their ejected mass in a short duration of time, 10$^{2}$$-$10$^{3}$ years.

\placetable{model_results}

A kinematical age since the end of the end of the high mass-loss phase was calculated based on the inner radius of the cool dust shell and the expansion velocity; 
this ranges from 1000 to 3000 yr for five of the objects but is only 300 yr for IRAS 22574+6609.
These values are consistent with the evolutionary ages for post-AGB stars \cite{blo95b}.
The results suggest that IRAS 22574+6609 may be more massive than the other sources, based 
on its larger value of the ejected mass, the larger expansion velocity, and the apparently
shorter time scale of evolution.
The kinematical ages also provide a time scale for the formation of the carrier of the 21 $\mu$m feature.

These new results for the mass loss rate and mass of the ejected shell can be compared with other published values determined in other ways. Here we have confined ourselves to two different approaches that also involve either discrete denser inner shells or steep density laws.

Ueta and Meixner have recently modeled the mid-infrared emission from the dusty envelopes of two carbon-rich PPNs based on high-resolution images.  They used a model that had an extended AGB envelope combined with an equatorially-enhanced discrete shell ejected during a later superwind phase.  Both of these are sources with 21 $\mu$m emission, although they are among the brighter ones and were not included in our present study.  For IRAS 22272+5435, the values determined were $\dot M$  = 4 $\times$ 10$^{-6}$  M$_\odot~{\rm yr}^{-1}$ and $\Delta$t = 1500 years \citep{uet01}.  For IRAS 07134+1005 (HD 56126), for which their model was based on CO interferometric observations in addition to mid-infrared images, they determined $\dot M$  = 3 $\times$ 10$^{-5}$  M$_\odot~{\rm yr}^{-1}$ and $\Delta$t = 840 years \citep{mei04}.  These mass loss rate values are a factor of 10 to 100 lower than what we have determined here.

Hrivnak \& Beiging (2005) determined mass loss rates and shell masses based on their measurements of CO gas in several transitional states, which allowed them to sample the CO emission further out and closer in to the star.  The observations were modelled with a one-dimensional statistical equilibrium, radiative transfer code.  They did not use a discrete shell with inner and outer boundary, as we have done here, but they did use a steep density law ($\alpha$=3), which produces a dense inner shell.
Their values for the mass loss range from a few $\times$ 10$^{-4}$  M$_\odot~{\rm yr}^{-1}$ 
to a few $\times$ 10$^{-5}$  M$_\odot~{\rm yr}^{-1}$, a factor of 10 less than we find here.
For three objects in common, our values for the mass loss rate larger by factors of two to 100.

A comparison of our new model results for mass loss rates with two different studies reveals that our rates are a few to 100 times larger, but are confined to a thinner shell.  As mentioned above, our mass loss rates are not as certain as we would like because the shell widths are not well constrained and assumed to be thin.  
Similar model fits could likely be produced with somewhat thicker shells with approximately the same total mass.  
Future higher resolution (0.1$\arcsec$) imaging than is presently available in the mid-infrared will better  constrain these shell widths.
It is the total mass that is the better determined parameter from our models.
Indeed, comparing the dust masses determined in these studies with our values of 0.7$-$2.8 $\times$ 10$^{-3}$ M$_\sun$ ($\Delta$M/330 from Table  \ref{model_results}), one finds much better agreement.  Meixner et al. determined a similar value of 0.8 $\times$ 10$^{-3}$ M$_\sun$ for IRAS 07134+1005, while Ueta et al. determined a value 20 times smaller, 
0.03 $\times$ 10$^{-3}$ M$_\sun$ for IRAS 22272+5435.  If we compare with the Hrivnak \& Bieging mass loss values based on CO gas and adjust with our gas-to-dust ratio of 330, we find dust masses 2$-$4 times smaller than ours, 0.24 $\times$ 10$^{-3}$ M$_\sun$ and 0.42 $\times$ 10$^{-3}$ M$_\sun$ for IRAS 07134+1005 and IRAS 22272+5435, respectively.  However, if our gas-to-dust mass ratio is overestimated by a
factor of 2$-$4, then the agreement would be excellent.  Note that Meixner et al. determined a gas-to-dust mass ratio of 75 for IRAS 07134+1005.


\section{RESULTS AND CONCLUSIONS}

Our targets were selected with the goal of completing the sample of carbon-rich 
PPNs studied with mid-infrared spectroscopy.  
However, three additional carbon-rich PPNs were identified in the past few years 
that were not included \citep{rey04,rey07}.  
Therefore the sample of carbon-rich PPNs that have been observed 
in mid-infrared spectroscopy is complete except for these three objects.  
Below are listed results based primarily on the targets of this study, supplemented  
by more general results for the sample.

1. Three new 21 $\mu$m sources were discovered: IRAS
06530$-$0213, 07430+1115 and 19477+2401.
This brings to 16 the total of 21 $\mu$m sources among the PPNs \citep{hri08}. 
Improved spectra of this feature were obtained for four others.

2. The shape and central wavelength (20.1$\pm$0.1 $\mu$m) 
of the the 21 $\mu$m feature
agrees with that found previously for the brighter sources using
{\it ISO} spectra.

3. These spectra do not show the resolution of the 30 $\mu$m
feature into two features as was claimed based on {\it ISO} spectra.
The suggested separation apparently was caused by a bad detector
band in the {\it ISO} spectra and is not real.

4. The 11.3 $\mu$m AIB feature is seen in all of the sources.  The
feature has a relatively consistent peak wavelength of 11.33$\pm$0.03 $\mu$m.  
However, the presence of the features at 12.3 and 13.3 $\mu$m and their
peak wavelengths and shapes vary among the sources.

5. These results strengthen the correlation between the presence
of the 21 $\mu$m feature and evidence of carbon-rich nature of the
sources.  In particular, all of the 21 $\mu$m sources are
carbon-rich (C/O$>$1.0), possess molecular carbon absorption
features (C$_2$, C$_3$), and possess AIB features \citep{hri08}.

6. All carbon-rich PPNs that have been observed possess 
the 21 $\mu$m, 30 $\mu$m, and AIB emission features.  
An exception may be IRAS 01005+7910 (OBe), the PPN with the 
hottest carbon-rich central star, which may or may not 
possess a weak 21 $\mu$m feature.
We also note that IRAS 19500$-$1709 is unusual in possessing 
the 21 $\mu$m (weak) and 30 $\mu$m (strong) features but weak or few 
AIB features \citep{vol02}. 

7. The 21 $\mu$m feature is present in all (but perhaps one) of the Galactic carbon-rich PPNs but is absent or weak in extreme carbon stars and PNs.  This suggests that the carrier is a common part of the outflow during the PPN phase and that the radiation field is important to the presence of the feature.  We propose (or This supports the idea) that the 21 $\mu$m carrier is produced in the extreme mass loss near the end of the AGB phase, is excited during the PPN phase, and is then destroyed in the more extreme UV radiation field of the subsequent PN phase.


8. A narrow feature at 13.7 $\mu$m due to  C$_2$H$_2$ is observed in the 
spectra of five objects;  it is seen in one source in absorption, in three in emission, 
and possibly one in both emission and absorption. 
These represent the first observations of C$_2$H$_2$ emission in a 
post-AGB object.
C$_2$H$_2$ absorption is commonly seen as a strong feature in the preceding 
AGB stage of stellar evolution.  While most may be incorporated into larger 
molecules in the evolution to PN, at least some is still present in some region 
around the star in the PPN phase.  
Studies of C$_2$H$_2$ provide a potential diagnostic of the physical conditions 
in the circumstellar envelope and should be followed up with 
high-resolution spectroscopy.

9. At $\sim$15.8 $\mu$m  a broad ($\sim$1.3 $\mu$m), unidentified emission 
feature is seen in the {\it Spitzer} spectra of four sources.  This feature also appears
to be present in {\it ISO} spectra of several additional 21 $\mu$m
sources.  
It is particularly strong in the two {\it Spitzer} sources with the
relatively strongest 21 $\mu$m feature. 
This suggests a possible correlation between the two
features which, if confirmed, could be a valuable aid in the
identification of the 21 $\mu$m feature.

10. 1-D modeling of the SEDs indicates a dense cool (120$-$150 K) shell of 
dust around the stars and also low density hot (500$-$800 K)  
dust closer to the star.  
The mass-loss rates when the shells detached are high, ranging from a 
10$^{-4}$ M$_\odot~{\rm yr}^{-1}$ to a 6 $\times$ 10$^{-3}$ M$_\odot~{\rm yr}^{-1}$ 
but the highest rates were not sustained for long.
These result in shell masses of 0.2 to 1.0 M$_\odot$, 
ejected over intervals of 160 to 820 years.

11. The results of the modeling of the dust shells imply ages since the termination 
of extensive mass loss of 1000$-$3000 years for all of the sources except 
IRAS 22574+6609.  These ages for the expansion of the shell are consistent with 
the evolutionary ages of low-mass, post-AGB central stars \citep{blo95b}, as they should be.  
IRAS 22574+6609 appears to have only relatively recently left the AGB, $\sim$300 years ago.

\acknowledgments

We thank Bill Forrest, Dan Watson, and Ben Sargent for helpful
conversations on the reduction of the IRS data and choice of
peak-up stars.  
We appreciate the many helpful comments by the referee, Sacha Hony, 
which served to greatly improve the paper.
This work is based on observations made with the
{\it Spitzer Space Telescope}, which is operated by the Jet
Propulsion Laboratory, California Institute of Technology, under a
NASA contract. Support for this research was provided by NASA
through contract 1276197 issued by JPL/Caltech.  This publication
makes use of data products from the Two Micron All Sky Survey,
which is a joint project of the University of Massachusetts and
the Infrared Processing and Analysis Center/California Institute
of Technology, funded by the National Aeronautics and Space
Administration and the National Science Foundation.  This research
has made use of the SIMBAD database, operated at CDS, Strasbourg,
France, and NASA's Astrophysical Data System.

Facilities: Spitzer

\clearpage

\begin{table}
\begin{center}
\caption{{\it Spitzer} Observing Log \label{obs_log}}
\begin{tabular}{rrrrr}
\tableline\tableline IRAS ID &\multicolumn{2}{c}{\underline
{~~~~~~~~~~~~LH~~~~~~~~~~~~}}
&\multicolumn{2}{c}{\underline {~~~~~~~~~~~~SH~~~~~~~~~~~~~}}\\
\multicolumn{1}{c}{}& \multicolumn{1}{c}{Date}&\multicolumn{1}{c}{Time\tablenotemark{a}}& \multicolumn{1}{c}{Date}&\multicolumn{1}{c}{Time\tablenotemark{a}}\\

\tableline
05113+1347 & 2006 Mar 17 & 3 $\times$ 14 & 2004 Oct 03\tablenotemark{b} & 3 $\times$ 30 \\
05341+0852 & 2007 Apr 16 & 3 $\times$ 14 & 2004 Mar 22\tablenotemark{b} & 3 $\times$ 30 \\
06530$-$0213 & 2006 Apr 25 & 3 $\times$ 14 & 2006 Apr 25 & 3 $\times$ 30 \\
07430+1115 & 2007 Apr 27  & 3 $\times$ 14  &  2007 Apr 27  & 3 $\times$ 30  \\
19477+2401 & 2005 Oct 16 & 3 $\times$ 6 & 2004 May 13\tablenotemark{b} & 3 $\times$ 6 \\
22574+6609 & 2005 Oct 12 & 3 $\times$ 6 & 2004 Aug 10\tablenotemark{b} & 4 $\times$ 30 \\
23304+6147 & 2005 Dec 15 & 3 $\times$ 6 & 2005 Dec 15 & 3 $\times$ 30 \\
\tableline
\end{tabular}
\tablenotetext{a}{Number of scan cycles times the on-source
integration time in seconds. } \tablenotetext{b}{Observed in
GTO93, PI: D. Cruikshank. }
\end{center}
\end{table}

\clearpage

\begin{table}
\begin{center}
\caption{Target Positions (2000.0)\label{pos}}
\begin{tabular}{cccc}
\tableline\tableline
IRAS ID& 2MASS ID & RA & Decl. \\
\tableline
05113+1347    & 05140775$+$1350282 & 05:14:07.76 & $+$13:50:28.3 \\
05341+0852    & 05365506$+$0854087 & 05:36:55.06 & $+$08:54:08.7 \\
06530$-$0230  & 06553181$-$0217283 & 06:55:31.82 & $-$02:17:28.3 \\
07430+1115    & 07455139$+$1108196 & 07:45:51.39 & $+$11:08:19.6 \\
19477+2401    & 19495491$+$2408532 & 19:49:54.91 & $+$24:08:53.3 \\
22574+6609    & 22591835$+$6625482 & 22:59:18.36 & $+$66:25:48.2  \\
23304+6147    & 23324479$+$6203491 & 23:32:44.79 & $+$62:03:49.1  \\
\tableline
\end{tabular}
\end{center}
\end{table}

\clearpage

\begin{table}
\begin{center} \caption{21 $\mu$m Feature Profile and Relative
Strengths\label{21pro}}
\begin{tabular}{cccc}
\tableline\tableline
 & Relative &  & \\
 & Feature  & Width (FWZL)\tablenotemark{a} & $\lambda_{peak}$\tablenotemark{b} \\
IRAS ID& Strength & ($\mu$m) & ($\mu$m) \\
\tableline
05113+1347    & 0.49  & 5.4  & 20.2\\
05341+0852    & 0.35  & 5.4 & 20.1\\
06530$-$0230  & 1.12   & 6.0 & 20.1\\
07430+1115    & 0.21  & 4.6 & 20.2\\
22574+6609    & 0.37  & 5.7  & 20.1 \\
23304+6147    & 1.46  & 5.7  & 20.1 \\
\tableline
\end{tabular}
\tablenotetext{a}{Full width at zero level. }
\tablenotetext{b}{Wavelength of feature peak in a plot of
F$_{\nu}$ vs. $\lambda$ . }
\end{center}
\end{table}

\clearpage

\begin{deluxetable}{rlclcccccl}
\rotate \tablecolumns{10} \tabletypesize{\scriptsize}
\tablewidth{9.6truein}
\tablecaption{Summary of Spectral Emission Features
\label{features}} \tablehead{
\colhead{IRAS ID} & \colhead{SpT} & \colhead{C/O\tablenotemark{a}} & \colhead{AIB Features (3$-$9 $\mu$m)\tablenotemark{b,c}} & \colhead{Ref\tablenotemark{d}} & \colhead{11.3 $\mu$m} & \colhead{21 $\mu$m} & \colhead{30 $\mu$m} & \colhead{13.7 $\mu$m\tablenotemark{e}} & \colhead{Other Features in Spitzer Spectra\tablenotemark{b}}  }\startdata
05113+1347    &G8 Ia   &2.4   &3.3:, 3.4: & 1     &Y  &Y &Y & N & 12.1, 13.3: 15.8(mw):, 22.3 \\
05341+0852    &G2 0-Ia &1.6   &3.3, 3.4, 6.2, 6.9, 8(br) & 1,2     &Y  &Y &Y & E & 12.4, 15.8(mw), (no LH {\it Spitzer} spectrum) \\
06530$-$0213  &F5 I    &2.8   &3.3        & 1     &Y  &Y &Y & E & 12.3, 13.2, 15.8(mw), 22.3 \\
07430+1115  &G5 0-Ia   &\nodata   &3.3, 3.4  & 1    &Y  &Y &Y & E: & 12(mw) \\
19477+2401    &\nodata\tablenotemark{g} &\nodata\tablenotemark{g}   &\nodata\tablenotemark{f} & 1 & \nodata\tablenotemark{h} &Y &Y & \nodata\tablenotemark{h} & (no SH spectrum), 22.3, 28.6:, 29.8:, 32.6: \\
22574+6609    &\nodata\tablenotemark{g} &\nodata\tablenotemark{g}   &\nodata\tablenotemark{f}, 6.2, 6.9, 7.7, 8(br) & 1,2  &Y  &Y &Y & A & 12.6, 17(mw) \\
23304+6147    &G2 Ia   &2.8   &3.3, 3.4:, 6.2, 6.9, 7.7, 8.6 (8br) & 1,3 &Y  &Y &Y & A/E: & 12.3, 13.2 15.9(mw), 22.3 \\
\enddata
\tablenotetext{a}{References to C/O abundances: Multiple objects -
\citet{win00}, \citet{red02}; IRAS 06530$-$0213 - \cite{rey04}. }
\tablenotetext{b}{Colons indicate an uncertain detection; mw =
medium-wide feature; abs = absorption feature. }
\tablenotetext{c}{Included also are aliphatic infrared bands at
3.4 and 6.9 $\mu$m. } \tablenotetext{d}{References for AIB
features: (1) \citet{hri07}; (2) \citet{hri00}; (3) review of ISO
archival spectrum. } \tablenotetext{e}{E = emission; A
= absorption feature; N = no feature. } 
\tablenotetext{f}{Not observed in the 3 $\mu$m spectral region. }
\tablenotetext{g}{Faint, with no visible spectra. } 
\tablenotetext{h}{Not observed in this spectral region with {\it Spitzer}. }
\end{deluxetable}

\clearpage

\thispagestyle{empty}
\setlength{\voffset}{25mm}
\begin{deluxetable}{cccccccccccccccccc}
\rotate \tablecolumns{18} \tabletypesize{\scriptsize}
\tablewidth{0truein}\center
\tablecaption{Results of 1-D Model Fits to the Spectral Energy
Distributions\label{models}} \tablehead{
\colhead{} & \colhead{} & \colhead{} & \colhead{} & \colhead{} & \colhead{} && \multicolumn{4}{c}{Cool Dust} && \multicolumn{3}{c}{Hot Dust} && \colhead{} & \colhead{} \\
\cline{8-11} \cline{13-15} \\
\colhead{IRAS ID} & \colhead{$\tau_o$} & \colhead{$\tau_o$} & \colhead{$\alpha$} & \colhead{$T_*$} & \colhead{$L_*/D^2$} && \colhead{$T_d$(c)\tablenotemark{a}} & \colhead{$r_{in}(c)/D$} & \colhead{$r_{out}(c)/D$} & \colhead{$\dot M/VD$} && \colhead{$T_d$(h)\tablenotemark{a}} & \colhead{$r_{in}(h)/D$} & \colhead{$\dot M/VD$} && \colhead{$\rho(c)$/$\rho(h)$\tablenotemark{b}} & \colhead{$A_V$} \\
\colhead{} & \colhead{(11.22 $\mu$m)} & \colhead{(V)} & \colhead{} & \colhead{(K)} & \colhead{($L_\odot$/kpc$^2$)}  
&& \colhead{(K)} & \colhead{($10^{-3}$ pc/kpc)} & \colhead{(($10^{-3}$ pc/kpc))} & \colhead{($\frac{M_\odot {\rm yr}^{-1}}{{\rm km~s}^{-1} {\rm kpc}^{-1}}$)} 
&& \colhead{(K)} & \colhead{($10^{-3}$ pc/kpc)} & \colhead{($\frac{M_\odot {\rm yr}^{-1}}{{\rm km~s}^{-1} {\rm kpc}^{-1}}$)} && \colhead{} & \colhead{(ISM)} }\startdata
05113$+$1347 & 0.0085 & 1.2 & 2.0 & 5500 & 170 && 132 & 4.2 & 5.4       & 4.5 $\times$ $10^{-6}$  && 714      & 0.043   & 3.4 $\times$ $10^{-10}$ && 12,500   & 2.8\\
05341$+$0852 & 0.0067 & 0.94& 3.5 & 6500 & 126 && 146 & 2.5 & \nodata & 1.0 $\times$ $10^{-6}$  && 486      & 0.117    &  6.7 $\times$ $10^{-9}$ && 15,000   & 3.7 \\
06530$-$0213 & 0.0052 & 0.73 & 3.0 & 7000 & 280 && 130 & 5.4 & \nodata & 1.6 $\times$ $10^{-6}$  && \nodata & \nodata & \nodata                          && \nodata  & 5.0\\
07430$+$1115 & 0.018 &  2.5 & 2.0\tablenotemark{c} & 5500 & 187 && 139 & 2.8 & 3.3       & 1.1 $\times$ $10^{-5}$  && \nodata & \nodata & \nodata                          && \nodata  & 1.5\\
22574$+$6609 & 0.39   & 55   & 2.0 & 5500 & 145  && 148 & 1.0  & 1.6      & 3.0 $\times$ $10^{-5}$  && 765      & 0.029    & 4.9 $\times$ $10^{-9}$ &&   7700   & 3.0\tablenotemark{c}\\
23304$+$6147 & 0.030 & 4.2   & 2.5 & 5500 & 320  && 122 & 5.0  & 6.4    & 2.0 $\times$ $10^{-5}$  && 609      & 0.083    & 1.7 $\times$ $10^{-9}$ && 10,000   & 1.8\\
\enddata
\tablenotetext{a}{Dust temperature at r$_{\rm in}$. }
\tablenotetext{b}{Density ratio at r$_{\rm in}(c)$. }
\tablenotetext{c}{Fixed value. }
\end{deluxetable}

\clearpage
\setlength{\voffset}{0mm}

\begin{table}
\begin{center} \caption{Model Results for Mass Loss Rates, Shell Masses, Ejection Durations, and Kinematical Ages\label{model_results}}
\begin{tabular}{ccccrcrr}
\tableline\tableline
 & D\tablenotemark{a}  & V\tablenotemark{b} & $\dot M$ & r$_{\rm in}$(c) & $\Delta$M & t($\Delta$M)\tablenotemark{c}& t$_{\rm kin}$\tablenotemark{d} \\
IRAS ID& (kpc) & (km~s$^{-1}$) & (M$_{\odot}$~yr$^{-1}$) & (pc) & (M$_{\odot}$) & (yr) & (yr) \\
\tableline
05113$+$1347 & 7.0 & 10 & 3.2 $\times$ $10^{-4}$ & 0.030 & 0.25 & 820 & 2900 \\
05341$+$0852 & 8.2 & 12 & 1.0 $\times$ $10^{-4}$ & 0.020 & 0.27 & \nodata & 1600 \\
06530$-$0230  & 5.4 & 14 & 1.2 $\times$ $10^{-4}$ & 0.030 & 0.22 & \nodata & 2100 \\
07430$+$1115 & 6.7 & 15 & 1.1 $\times$ $10^{-3}$ & 0.019 & 0.27 & 180 & 1200 \\
22574$+$6609 & 7.6 & 26 & 5.8 $\times$ $10^{-3}$ & 0.008  & 0.95 & 160 &  280 \\
23304$+$6147 & 5.1 & 12 & 1.2 $\times$ $10^{-3}$ & 0.025  & 0.66 & 580 & 2000 \\
\tableline
\end{tabular}
\tablenotetext{a}{Distance based upon assumed luminosity of 8300 L$_{\sun}$.}
\tablenotetext{b}{Expansion velocities as found in the literature and averaged in some cases.}
\tablenotetext{c}{Duration of enhanced mass loss, determined from (r$_{\rm out}$(c) - r$_{\rm in}$(c))/V. }
\tablenotetext{d}{Kinematical age, determined from r$_{\rm in}$(c)/V. }
\end{center}
\end{table}

\clearpage

\clearpage

\clearpage

\clearpage

\clearpage

\clearpage

\begin{thebibliography}{}

\bibitem[Bakker et al.(1997)]{bak97}
    Bakker, E. J., van Dishoeck, E. F., Waters, L. B. F. M.,
    \& Schoenmaker, T. 1997, \aap, 323, 469
\bibitem[Bernard-Salas et al.(2006)]{ber06}
    Bernard-Salas, J., Peeters, E., Sloan, G.C., Cami, J., Guiles, S., \& Houck, J.R.
    2006, \apjl, 652, L29
\bibitem[Bl\"{o}cker(1995a)]{blo95a}
    Bl\"{o}cker, T. 1995a, \aap, 297, 755
\bibitem[Bl\"{o}cker(1995b)]{blo95b}
    Bl\"{o}cker, T. 1995b, \aap, 299, 727
\bibitem[Buchanan et al.(2006)]{buc06}
    Buchanan, C. L., Kastner, J. H, Forrest, W.J., et al. 2006, \aj, 132,
    1890
\bibitem[Cernicharo et al.(2001)]{cer01}
    Cernicharo, J., Heras, A. M., Tielens, A. G. G. M., et al.
    2001, \apj, 546, L123
\bibitem[Cernicharo et al.(1999)]{cer99}
    Cernicharo, J., Yamamura, I., Gonz\'{a}lez-Alfonso, E., et al. 
    1999, \apj, 526, L41
\bibitem[Forrest, Houck, \& McCarthy(1981)]{for81}
    Forrest, W. J., Houck, J. R., \& McCarthy, J. F. 1981, \apj, 248,
    195
\bibitem[Garc\'{i}a-Hern\'{a}ndez et al.(2002)]{garher02}
    Garc\'{i}a-Hern\'{a}ndez, D.A., Manchado, A., Garc\'{i}a-Lario, P.,
    Dom\'{i}ngueq-Tagle, C., Conway, G.M., \& Prada, F. 2002, \aap, 387,
    955
\bibitem[Garc\'{i}a-Lario et al.(1999)]{garlar99}
    Garc\'{i}a-Lario, P, Manchado, A., Ulla, A., \& Manteiga, M. 1999, 
    \apj, 513, 941
\bibitem[Goebel \& Moseley(1985)]{goe85}
    Goebel, J. H., \& Moseley, S. H. 1985, \apj, 290, L35
\bibitem[Higdon et al.(2004)]{hig04}
    Higdon, S. J. U., et al. 2004, \pasp, 116, 975
\bibitem[Hony et al.(2001)]{hon01a}
    Hony, S., Van Kerckhoven, C., Peeters, E., Tielens, A. G. G. M., 
    Hudgins, D.M., \& Allamandola, L.J. 2001, \aap, 370, 1030
\bibitem[Hony, Waters, \& Tielens(2001)]{hon01}
    Hony, S., Waters, L. B., F. M., \& Tielens, A. G. G. M. 2001, \aap, 378,
    L41
\bibitem[Hony, Waters, \& Tielens(2002)]{hon02}
    Hony, S., Waters, L. B., F. M., \& Tielens, A. G. G. M. 2002, \aap, 390,
    533
\bibitem[Houck et al.(2004)]{hou04}
    Houck, J. R., et al. 2004, \apjs, 154, 18
\bibitem[Hrivnak(1995)]{hri95}
    Hrivnak, B.J. 1995, \apj, 438, 341
\bibitem[Hrivnak, Geballe, \& Kwok(2007)]{hri07}
    Hrivnak, B.J., Geballe, T. R., \& Kwok, S. 2007, \apj, 662,
    1059
\bibitem[Hrivnak \& Kwok(1991b)]{hri91b}
    Hrivnak, B.J., \& Kwok, S. 1991b, \apj, 368, 564
\bibitem[Hrivnak \& Kwok(1991a)]{hri91a}
    Hrivnak, B.J., \& Kwok, S. 1991a, \apj, 371, 631
\bibitem[Hrivnak \& Kwok(1999)]{hri99}
    Hrivnak, B.J., \& Kwok, S. 1999, \apj, 513, 869
\bibitem[Hrivnak et al.(1985)]{hri85}
    Hrivnak, B.J., Kwok, S., \& Boreiko, R.T.  1985, \apjl, 294,
    L113
\bibitem[Hrivnak et al.(2008)]{hri08}
    Hrivnak, B.J., Volk, K., Geballe, T.R., \& Kwok, S. 2008, 
    in IAU Symp. 251: Organic Matter in Space, eds. S. Kwok,
    S.A. Sanford, in press
\bibitem[Hrivnak, Volk, \& Kwok(2000)]{hri00}
    Hrivnak, B.J., Volk, K., \& Kwok, S. 2000, \apj, 535, 275
\bibitem[Hudgins \& Allamandola(1999)]{hud99}
    Hudgins, D.M., \& Allamandola, L.J. 1999, \apjl, 516, L41
\bibitem[Jiang et al.(2005)]{jia05}
    Jiang, B.W., Zhang, K., \& Li, A. 2005, \apjl, 630,
    L77
\bibitem[Keady \& Ridgway(1993)]{kea93}
    Keady, J.J., \& Ridgway, S.T. 1993, \apj, 406, 199
\bibitem[Kelly \& Hrivnak(2005)]{kel05}
    Kelly, D. M., \&  Hrivnak, B. J. 2005, \apj, 629, 1040
\bibitem[Kimura et al.(2005)]{kim05}
    Kimura, Y., Nuth, J.A., III, \& Ferguson, F.T. 2005, \apjl, 632,
    L159
\bibitem[Kraemer et al.(2006)]{kra06}
    Kraemer, K., Sloan, G.C., Bernard-Salas, J., Price, S.D., 
    Egan, M.P., \& Wood, P.R. 2006, \apjl, 652, L25
\bibitem[Kurucz(1993)]{kur93}
    Kurucz, R. L. 1993, ``ATLAS9 Stellar Atmosphere Programs and 2 km/s Grid,''
	Kurucz CD-ROM No. 13. Cambridge, Mass.: Smithsonian Astrophysical 
	Observatory
\bibitem[Kwok, Hrivnak, \& Geballe(1995)]{kwo95}
    Kwok, S., Hrivnak, B.J., \& Geballe, T. R. 1995, \apj, 454, 394
\bibitem[Kwok, Volk, \& Bernath(2001)]{kwo01}
    Kwok, S., Volk, K., \& Bernath, P. 2001, \apj, 554, L87
\bibitem[Kwok, Volk, \& Hrivnak(1989)]{kwo89}
    Kwok, S., Volk, K., \& Hrivnak, B.J. 1989, \apj, 345, L51
\bibitem[Kwok, Volk, \& Hrivnak(1999a)]{kwo99a}
    Kwok, S., Volk, K., \& Hrivnak, B.J. 1999a, in IAU Symp. 191: Asymptotic
    Giant Branch Stars, ed. T. Le Bertre, A. L\`{e}bre, C. Waelkens (San Francisco:
    ASP), 297
\bibitem[Kwok, Volk, \& Hrivnak(1999b)]{kwo99b}
    Kwok, S., Volk, K., \& Hrivnak, B.J. 1999b, \aap, 350, L35
\bibitem[Leung(1976)]{leu76}
    Leung, C.M. 1976, \apj, 209, 75
\bibitem[Men'shchikov, Hofmann, \& Weigelt(2002)]{men02}
    Men'shchikov, A.B., Hofmann, K.-H., \& Weigelt, G. 2002, \aap,
    392, 921
\bibitem[Meixner et al.(2004)]{mei04} 
	Meixner, M., Zalucha, Z., Ueta, T., Fong, D., \& Justtanont, K. 2004, \apj, 614, 371
\bibitem[Moutou et al.(1999)]{mou99}
    Moutou, C., Sellgren, K., L'{e}ger, A., Verstraete, L., \& Le Coupanec, P. 
    1999, in Solid Interstellar Matter: The ISO Revolution, ed. L. d'Hendecourt, 
    C. Joblin, A. Jones (Springer-Verlag: Berlin), 89
\bibitem[Neckel \& Klare(1980)]{nic80}
    Neckel, Th., \& Klare, G 1980, \aaps, 42, 251
\bibitem[Omont et al.(1995)]{omo95}
    Omont, A., Moseley, S. H., Cox, P., et al. 1995, \apj, 454,
    819
5    Oppenheimer, B. D., Bieging, J. H., Schmidt, G.D., Gordon, K.D., Misselt, K.A., 
\bibitem[Peeters et al.(2002)]{pee02}
    Peeters, E., Hony, S., Van Kerckhoven, C., Tielens, A.G.G.M., 
    Allamandola, L.J., Hudgins, D.M., \& Bauschlicher, C.W. 2002,
     \aap, 390, 1089
\bibitem[Posch et al.(2004)]{pos04}
    Posch, Th., Mutschke, H., \& Andersen, A. 2004, \apj, 616,
    1167
\bibitem[Reddy et al.(2002)]{red02}
    Reddy, B. E., Lambert, D. L., Gonzalez, G., \& Yong, D. 2002, \apj, 564,
    482
\bibitem[Reyniers et al.(2007)]{rey07}
    Reyniers, M., Van de Steene, G.C., van Hoof, P.A.M., \& Van Winckel, H.
    2007, \aap, 471, 247
\bibitem[Reyniers et al.(2004)]{rey04}
    Reyniers, M., Van Winckel, H., Gallino, R., \& Straniero, O. 2004, \aap, 417,
    269
\bibitem[Rouleau \& Martin(1991)]{rou91}
    Rouleau, F., \& Martin, P.G. 1991, \apj, 377, 526
\bibitem[Sahai et al.(2007)]{sah07}Sahai, R., Morris, M., S\'{a}nchez Contreras, C.,
    \& Claussen, M. 2007, \aj, 134, 2200
\bibitem[S\'{a}nchez Contreras et al.(2006)]{san06}
    S\'{a}nchez Contreras, C., Le Mignant, D., Sahai, R., Chaffee, F.H., \&
    Morris, M. 2006, in IAU Symp. 234: Planetary Nebulae in our Galaxy and Beyond, 
    ed. M.J. Barlow, R.H. M\'{e}ndez (Cambridge: Cambridge University Press), 71
\bibitem[Sellgren, Uchida, \& Werner(2007)]{sel07}
    Sellgren, K., Uchida, K.I., \& Werner, M.W. 2007, \apj, 659, 1338
\bibitem[Sloan et al.(2007)]{slo07}
    Sloan, G.C., Jura, M., Duley, W.W., et al. 2007, \apj, 664, 1144
\bibitem[Sloan et al.(2005)]{slo05}
    Sloan, G.C., Keller, L.D., Forrest, W.J., et al. 2005, \apj, 632, 956
\bibitem[Speck \& Hofmeister(2004)]{spe04}
    Speck, A. K. \& Hofmeister, A. M. 2004, \apj, 600, 986
\bibitem[Su et al.(2001)]{su01}
    Su, K.Y.L., Hrivnak, B.J., \& Kwok, S. 2001, \aj, 122, 1525
\bibitem[Szczerba et al.(1999)]{szc99}
    Szczerba, R., Henning, Th., Volk, K., Kwok, S., \& Cox, P. 1999, \aap,
    345, L39
\bibitem[Ueta et al.(2000)]{uet00} 
	Ueta, T., Meixner, M., \& Bobrowsky, M. 2000, \apj, 528, 861
\bibitem[Ueta et al.(2001)]{uet01} 
	Ueta, T., Meixner, M., Hintz, P.M., et al. 2001, \apj, 557, 831
\bibitem[van Diedenhoven et al.(2004)]{vandie04}
    van Diedenhoven, B.,  Peeters, Van Kerckhoven, C.,E., Hony, S., 
    Hudgins, D.M., Allamandola, L.J., \& Tielens, A. G. G. M., 
    2004, \apj, 611, 928
\bibitem[Van Kerckhoven et al(2000)]{vanker00}
    Van Kerckhoven et al. 2000, \aap, 357, 1013
\bibitem[Van Winckel(2003)]{vanwin03}
    Van Winckel, H. 2003, \araa, 41, 397
\bibitem[Van Winckel \& Reyniers(2000)]{win00}
    Van Winckel, H. \& Reyniers, M. 2000, \aap, 354, 135
\bibitem[Volk (2003)]{vol03}
    Volk, K., 2003,  in IAU Symp. 209: Planetary Nebulae $-$ Their Evolution and Role
    in the Universe, ed. S. Kwok, M. Dopita, R. Sutherland (San Francisco: ASP), 281
\bibitem[Volk \& Kwok(1989)]{vol89}
    Volk, K., \& Kwok, S. 1989, \apj, 342, 345
\bibitem[Volk, Kwok, \& Hrivnak(1999)]{vol99}
    Volk, K., Kwok, S., \& Hrivnak, B.J. 1999, \apj, 516, L99
\bibitem[Volk et al.(2002)]{vol02}
    Volk, K., Kwok, S., Hrivnak, B.J., \& Szczerba, R. 2002, \apj, 567, 412
\bibitem[Volk, Xiong, \& Kwok (2000)]{vol00}
    Volk, K., Xiong, G. Z., \& Kwok, S. 2000, \apj, 530, 408
\bibitem[von Helden et al.(2000)]{von00}
    von Helden, G., et al. 2000, Science, 288, 313
\bibitem[Werner et al.(2004)]{wer04}
    Werner, M., et al. 2004, \apjs, 154, 11
\bibitem[Willems(1988)]{wil88}
    Willems, F.J.1988, \aap, 203, 51
\bibitem[Zhukovska \& Gail (2008)]{zhu08}
    Zhukovska, S., \& Gail, H.-P. 2008, \aap, 486, 229 

\end{thebibliography}
\end{document}